\documentclass[fleqn,useAMS,usenatbib]{mnras}
\usepackage{times}
\usepackage{rotating}
\usepackage{amsmath}
\usepackage{amssymb}
\usepackage[dvipsnames]{xcolor}
\usepackage{graphicx}
\usepackage{graphics}
\usepackage{hyperref}
\hypersetup{
	colorlinks   = true,
	citecolor    = blue
}

\def\apj{{ ApJ}}
\def\apjl{{ApJL}}
\def\apjs{{ ApJS}}

\def\aap{{ A\&A}}
\def\aj{{ AJ}}
\def\mnras{{ MNRAS}}
\def\araa{{ ARA\&A}}

\def\nat {{ Nature}}

\def\physrep{{ Physics Reports}}

\def\prd{{ Phys. Rev. D}}

\def\pasa{{PASA}}

\long\def\symbolfootnote[#1]#2{\begingroup%
	\def\thefootnote{\fnsymbol{footnote}}\footnote[#1]{#2}\endgroup}

\newcommand{\gae}{\lower 2pt \hbox{$\, \buildrel {\scriptstyle >}\over {\scriptstyle \sim}\,$}}
\newcommand{\lae}{\lower 2pt \hbox{$\, \buildrel {\scriptstyle <}\over {\scriptstyle \sim}\,$}}
\newcommand{\aprop}{\lower 2pt \hbox{$\, \buildrel {\scriptstyle \propto}\over {\scriptstyle \sim}\,$}}

\newcommand\restr[2]{{
		\left.\kern-\nulldelimiterspace 
		#1 
		\vphantom{\big|} 
		\right|_{#2} 
}}
\def\d{\mathrm{d}}

\def\n13{\mbox{\large n}_{_{13}}}

\begin{document}
	
	\title[H reionization with FRBs]
	{Exploring the epoch of hydrogen reionization using FRBs}
	
	\author[Beniamini et al.]{Paz Beniamini$^{1,2}$, Pawan Kumar$^3$, Xiangcheng Ma$^4$, Eliot Quataert$^{4,5}$
		\\ $^1$Division of Physics, Mathematics and Astronomy, California Institute of Technology, Pasadena, CA 91125, USA \\
		$^2$Astrophysics Research Center of the Open University (ARCO), The Open University of Israel, P.O Box 808, Ra’anana 43537, Israel\\
		$^{3}$Department of Astronomy, University of Texas at Austin, Austin, TX 78712, USA\\
		$^{4}$Department of Astronomy and Theoretical Astrophysics Center, University of California Berkeley, Berkeley, CA 94720, USA\\
		$^{5}$Department of Astrophysical Sciences, Princeton University, Princeton, NJ 08544, USA}
	
	\maketitle
	
	\begin{abstract}
		We describe three different methods for exploring the hydrogen reionization epoch using fast radio bursts (FRBs) and provide arguments for the existence of FRBs at high redshift (z). The simplest way, observationally, is to determine the maximum dispersion measure (DM$_{\rm max}$) of FRBs for an ensemble that includes bursts during the reionization. The DM$_{\rm max}$ provides information regarding reionization much like the optical depth of the CMB to Thomson scattering does, and it has the potential to be more accurate than constraints from Planck, if DM$_{\rm max}$ can be measured to a precision better than 500 $\mbox{pc cm}^{-3}$. Another method is to measure redshifts of about 40 FRBs between z of 6-10 with$\sim10\%$ accuracy to obtain the average electron density in 4 different z-bins with $\sim4\%$ accuracy. These two methods don't require knowledge of the FRB luminosity function and its possible redshift evolution. Finally, we show that the reionization history is reflected in the number of FRBs per unit DM, given a fluence limited survey of FRBs that includes bursts during the reionization epoch; we show using FIRE simulations that the contributions to DM from the FRB host galaxy \& CGM during the reionization era is a small fraction of the observed DM. This third method requires no redshift information but does require knowledge of the FRB luminosity function. 
	\end{abstract}
	

	\section{Introduction}
	
	Fast Radio Bursts (FRBs) are bright, few ms duration, radio pulses that have been detected between about $400$\,MHz and $7$\,GHz. Their rate is estimated to be $\sim 10^4$ per day. These bursts are widely distributed with many located to galaxies at distances of a few Gpc \citep{Lorimer+07,Thornton+13,Spitler2014,Petroff2016,Bannister2017,Law+17,Chatterjee+17,Marcote2017,Tendulkar+17,Gajjar2018,Michilli+18,Farah2018,Shannon2018,Oslowski2019,Kocz2019,Bannister+19,CHIME2019,CHIME2019b,Ravi2019,Ravi2019b,Ravi+19}. 
	
	
	The arrival time of the signal at different frequencies gives us the column density of electrons between a burst and us (dispersion measure - DM). At sufficiently large distances, the latter are dominated by propagation through the intergalactic medium (IGM). As a result, the distance to FRBs can be roughly estimated from the DM. The largest measured value of DM for a FRB to date is 2500$\mbox{pc cm}^{-3}$, which corresponds to a redshift $z\sim 3$ \citep{Zhang2018}.
	The accurate measurement of FRBs DMs makes them attractive probes for determining the baryon content of the universe e.g. \citep{Macquart20}, exploring the helium reionization epoch, e.g. \citep{Caleb2019,Linder2020,Bhattacharya2020}, and constraining a number of other cosmological properties \citep{Deng2014,YZ2017,Walters2018,Jaroszynski2019,Wucknitz2020}. We investigate in this work whether they could also be used as a probe of the hydrogen reionization epoch.

	The discovery of an FRB in the Galaxy associated with a magnetar (\citealt{STARE2020,CHIME2020}) suggests that at least some, and perhaps most, FRBs are produced by magnetars. In this case, there should be a large number of FRBs from when the Universe was about 500 million years old and was undergoing reionization (a timescale much longer than the lifespan of massive stars that leave behind neutron star (NS) remnants, which is $\sim5-30$\,Myr). FRBs should exist at $z>6$ for several reasons. It is widely believed that UV radiation from massive stars ($M \gae 10 M_\odot$) is responsible for reionizing the universe at $z>6$, and stars with mass between 
	$\sim10$ \& 40 $M_\odot$ leave behind NS remnants, e.g. \citep{Muno2006}. If the NSs' magnetic field is generated by dynamo (e.g. \citealt{TD1993}) within a few ms of their birth, then we expect a fraction of these NSs to have magnetar strength magnetic field, i.e. surface field larger than $10^{14}$\,G. At $z=0$, the fraction of NSs born as magnetars (as inferred from systems in our Galaxy) is estimated to be quite large, $\sim 40$\% \citep{Beniamini2019}. The case for magnetar-strength field for high-$z$ NSs is strengthened as they are likely born rotating faster\footnote{High-$z$, lower metalicity stars should be rotating faster at time of core collapse as they have smaller mass loss and retain a larger fraction of their angular momentum.} and should have large scale field generation via $\alpha$--$\Omega$ dynamo \cite{TD1993}. A useful comparison can also be made with gamma-ray bursts (GRBs), that have also been suggested as potential probes of H reionization \citep{Ioka2003,Inoue2004}. Empirically we know that there are many GRBs at $z > 6$ -- the largest measured $z$ for a GRB is 9.4 --  and the generation of relativistic jets in bursts almost certainly requires strong magnetic field in the cores of massive stars \citep{KZ2015}. Thus, there is a mechanism that generates magnetic fields in stars even at $z\sim 10$. Although, this is an indirect argument for the existence of magnetars at high $z$, it is tied to what we already know Nature can do, and its extrapolation to NSs is not a big stretch.

	Gunn-Peterson troughs \citep{GP1965} in the spectra of quasars at $z>6$, e.g. \cite{Fan2006}, suggest that the reionization of IGM hydrogen was largely completed by $z\sim 6$. Moreover, good constraints have been obtained regarding the mid-point of the cosmic reionization redshift by accurate measurement of the Thomson optical depth of the CMB by the Planck satellite from the measurement of large scale polarization anisotropies. Their most recent analysis finds the optical depth of microwave photons to Thomson scattering by free electrons in the Universe to be $\tau_{\rm T} =  0.0544\pm 0.0073$ \citep{Planck2020}. This optical depth translates to a reionization redshift of $z = 7.7\pm 0.8$ if the reionization were to be instantaneous, which of course is unphysical. For a physically more reasonable scenario, one might conclude, given the value of $\tau_{\rm T}$ measured by Planck, that the IGM was $\sim 50$\% neutral between  $z\sim 7$ \& 8.5 (see \citealt{Paoletti2020} for a more detailed approach). Since $\tau_{\rm T}$ is an integral of redshift weighted electron density ($n_{\rm e}$), it does not tell us how the reionization progressed with $z$, i.e. an infinite number of different $n_{\rm e}(z)$ functions give the same $\tau_{\rm T}$.
	
	The resonant scattering of Lyman-$\alpha$ photons by neutral hydrogen has a large cross-section, and spectroscopic studies of high-$z$ galaxies in Lyman-$\alpha$ provide a probe of the hydrogen reionization time-line. The number of galaxies with Lyman-$\alpha$ emission decreases rapidly for $z>6$, and the high redshift observations suggests that the IGM was $\sim 70$\% neutral between $z\sim7$ \& 8, e.g. \citep{Robertson2015,Dayal2018,Mason2018,Mason2019,Finkelstein2019,Hoag2019,Whitler2020} as the majority of $z>7$ galaxies were not detected spectroscopically. In principle, an accurate measurement of Lyman-$\alpha$ emission lines and their equivalent widths for a large number of galaxies can map the epoch of reionization. In practice, these measurements are sensitive to the neutral hydrogen column density and covering fraction in the host galaxy as well as in the IGM, and that can introduce systematic biases and uncertainties.
	
	As mentioned above, electron column densities, or DMs, are accurately measured for FRBs we detect. We explore here what it would take to turn the DM measurements for a set of FRBs to teach us about the reionization history. If FRBs can be used for this purpose, they offer a new way of exploring this important epoch and they are subject to a different set of systematic and random errors than the existing methods we have currently available. Furthermore, even a small subset of high-$z$ FRBs (most likely the repeaters), those that can be localized accurately and their redshifts measured from follow-up optical/IR observations, offer important additional information regarding reionization of the IGM. The reason is that these FRBs with measured $z$, give us an effective $\tau_{\rm T}$ (modulo the redshift weight factor) for different distances and lines-of-sight through the IGM. We investigate in this work what more we can learn from these measurements than we have from a single $\tau_{\rm T}$ value obtained for the CMB.

	If one were to assume that the reionization history of the universe is known from some other measurements -- Ly-$\alpha$ emission from galaxies for instance -- then that information can be used to convert DM for FRBs to a rough redshift measurement. This is because the contributions to the DM from the FRB host galaxy and circumgalactic medium (CGM) is relatively small (see \S\ref{fire-DM}), and the contribution from our galaxy can be subtracted reasonably well. We investigate what the FRB redshift determined in this way might teach us about cosmological parameters.

	\section{FRB DM distribution and reionization: a simplified analytic model}
	\label{sec:FRBdistAna}
	\subsection{Basics}
	
\begin{figure}
		\centering
		\includegraphics[width = 0.4\textwidth]{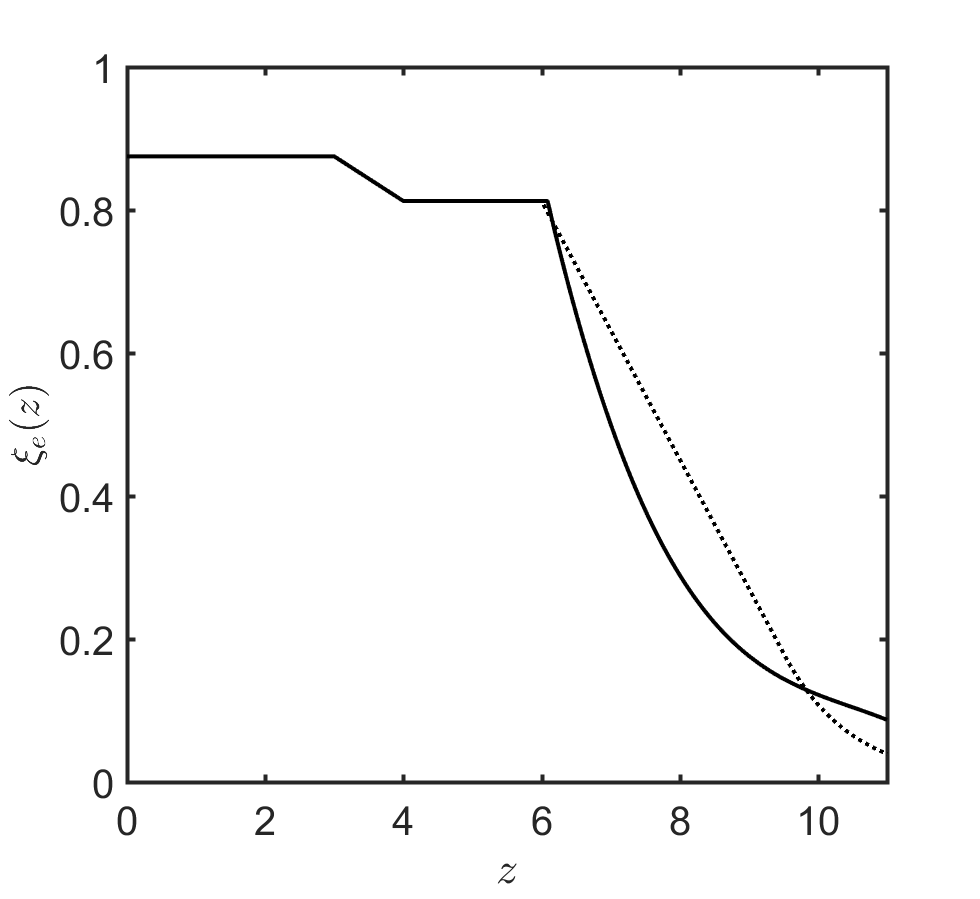}\\
		\includegraphics[width = 0.4\textwidth]{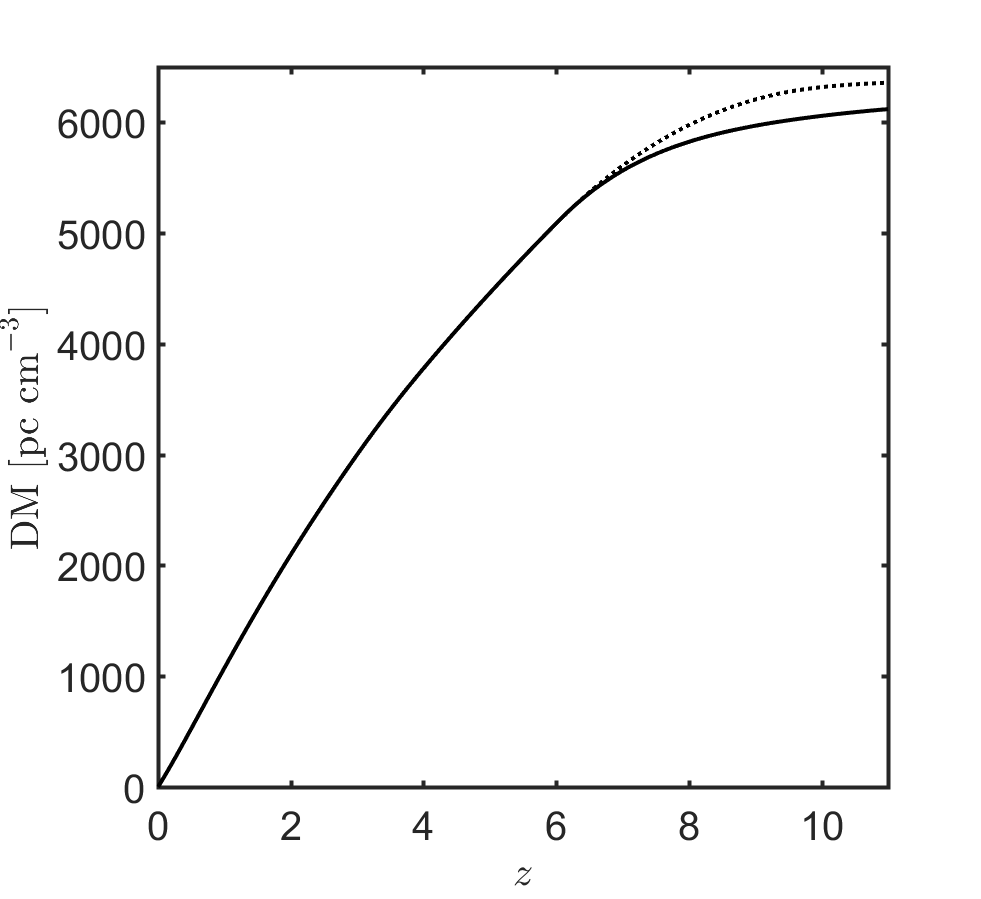}
		\vskip -0.2cm
		\caption{The upper panel shows the number of electrons per baryon, $\xi_e$, as a function of redshift for two different H-reionization models. The solid line represents the current observational estimates for $\xi_e$ at redshift larger than 6, cf. \citep{Finkelstein2019,Robertson2015} which we refer to as $\xi_{\rm e,o}(z)$. The dotted curve (corresponding to $\xi_{\rm e,t}(z)$ which is given by Eq. \ref{xi-test}) is a model we made up as a combination of linear and exponential functions to determine whether FRBs can discriminate between different reionization histories. The lower panel shows dispersion measure (DM) as a function of $z$ for these two different hydrogen reionization histories.}
		\label{fig:dm-z}
	\end{figure}

	The arrival time of a signal at frequency $\nu_{\rm obs}$ in the observer frame (relative to the moment of arrival of a signal traveling at $c$ from the same source) is given by
	\begin{equation}
	t = \int d\ell' {(1+z) \nu_p^2\over 2 c\nu^2} = 
	\int_0^r dr'\, a (1+z) {q^2 n_{\rm e} \over 2\pi m_{\rm e} c\nu^2}, 
	\end{equation}
	where $d\ell' = a dr'$ is the proper length of a small segment along the photon
	trajectory whereas $dr$ is the comoving length of the segment, $a=(1+z)^{-1}$ 
	is the scale factor of the universe, $n_{\rm e}$ is the proper density of electrons, 
	$\nu = \nu_{\rm obs} (1+z)$ is comoving photon frequency, $\nu_p^2=q^2 n_{\rm e}/\pi m_{\rm e}$ 
	the plasma frequency squared, $q$ \& $m_{\rm e}$ are electron charge and mass, and $c$ the speed of light. The arrival time of the signal at $\nu_{\rm obs}$ can be expressed as the integral of electron density or dispersion measure (DM)
	times a constant factor. Where
	\begin{equation}
	\mbox{DM} \equiv \int_0^r dr'\, {a\, n_{\rm e} \over 1+z'} = c \int_0^z dz' {n_{\rm e}(z') \over 
		(1+z')^2 H(z')}
	\label{dm-z1}
	\end{equation}
	$H(z) = d\ln a/dt$ is the Hubble constant, which can be expressed in terms
	of the critical density $\rho_{cr}$ for a flat universe: $H^2 = 8\pi 
	G\rho_{cr}/3$. Writing the density as a sum of matter and dark energy
	components, we arrive at
	\begin{equation}
	H^2 = {8\pi G (\rho_m + \Lambda)\over 3} 
	= H_0^2 \left[\Omega_{m0} (1+z)^3 + \Omega_{\Lambda0}\right].
	\end{equation}
	where $\Omega_{m0}\equiv \rho_{m}(z=0)/\rho_{cr}(z=0)$ is the dimensionless matter density and $H_0$ is the Hubble constant at $z=0$. In what follows we adopt a standard flat $\Lambda$CDM cosmology with parameters $H_0=68\mbox{km s}^{-1}\mbox{ kpc}^{-1}, \Omega_{\Lambda0}=0.69, \Omega_{m0}=1-\Omega_{\Lambda0}=0.31,\Omega_{b0}=0.048$ \citep{Planck2016}. 
	Defining the free electron fraction per nucleon, $\xi_e(z) = m_p {\rm n_{\rm e}}/\rho_b$ (where $\rho_b$ is comoving baryon density) and combining the above equations we find
	\begin{equation}
	\mbox{DM} = {3c H_0\Omega_{bo}\over 8\pi G m_p} \int_0^z dz' \, 
	{(1+z)\xi_e(z)\over \left[ (\Omega_{m0} (1+z)^3 + \Omega_{\Lambda0} 
		\right]^{1/2} }.
	\label{dm-z2}
	\end{equation}
	We assume here that H reionization is approximately concurrent with the ionization of HeI to HeII (see e.g. \citealt{Eide2020}) and that the He mass fraction is $Y\!=\!0.25$. When hydrogen is completely ionized and helium is singly ionized, $\xi_e = 0.8125$, and $\xi_e = 0$ when they are both entirely in atomic form. At a lower redshift, $3\lesssim z\lesssim 4$, helium becomes fully ionized (HeII to HeIII), and then we have $\xi_e = 0.875$. For simplicity we have assumed here that none of the gas is incorporated back into stars. This is a safe assumption when considering high redshift FRBs, during the H reionization epoch, as we do in this paper, and will only lead to a small error in their absolute DM values.
	We show in Fig. \ref{fig:dm-z} $\xi_{\rm e}$ 
	and the $\mbox{DM}$ as a function of $z$ for two different reionization scenarios. The first represents an estimate based on current observational constraints, adopted from \cite{Robertson2015}, hereby noted as $\xi_{\rm e,o}(z)$. The second is a simple `made up' model described by: 
	\begin{equation}
	\xi_{\rm e,t}(z) = \left\{
	\begin{array}{l}
	\hskip -5pt 0.875  \hskip 119pt z < 3  \\
	\hskip -5pt 0.875-0.0625 (z-3)  \hskip 55pt 3 < z < 4  \\   
	\hskip -5pt 0.8125  \hskip 115pt 4< z < 6  \\
	\hskip -5pt  0.8125(1-(z - 6)/4.5)    \hskip 48pt      6< z <9.5 \\
	\hskip -5pt  0.18\exp[-1.015(z-9.5)]  \hskip 41pt  z>9.5
	\end{array}
	\right.
	\label{xi-test}
	\end{equation}
	This expression for $\xi_e$ for $3 < z < 4$ approximately takes into account the second helium reionization and above $z=6$ accounts for the first helium reionization and the hydrogen reionization. Note that \cite{Robertson2015} provide ionization fraction for $z\gtrsim 6$. At lower redshifts we therefore adopt the same ionization histories for both $\xi_{\rm e,o}(z)$ and $\xi_{\rm e,t}(z)$. We stress that the results in this paper are largely independent of the assumptions regarding the details of the HeII to HeIII reionization, as we are primarily interested at the distribution of bursts at significantly higher redshifts/DMs.
	The purpose of the test model, $\xi_{\rm e,t}(z)$, is simply to demonstrate that the technique outlined in this paper has the capacity to differentiate between different hydrogen reionization evolutions. We have verified that the differences between $\xi_{\rm e,o}(z)$ and $\xi_{\rm e,t}(z)$ in terms of their affects on the DM distribution are typical. To do so, we have constructed a set of $10^4$ random $\xi_e(z)$ functions that are all chosen such that they are consistent with the determination of $\tau_{\rm T}$ using the Planck measurements (see \S \ref{sec:dmmax} for details). We found that the median difference in the maximum value of DM (described in \S \ref{sec:dmmax}) for a random pair of models is $210\mbox{ pc cm}^{-3}$, comparable to that between $\xi_{\rm e,o}(z)$ and $\xi_{\rm e,t}(z)$ ($240\mbox{ pc cm}^{-3}$). In addition, the difference in the ratio $N_{6000-7000}/N_{5000-6000}$ (see \S \ref{sec:MonteCarlospecific}) is 0.085 for a random pair of $\xi_e(z)$ functions as described above, as compared to 0.087 between $\xi_{\rm e,o}(z)$ and $\xi_{\rm e,t}(z)$.
	
	\subsection{FRB rate and their DM distribution}
	\subsubsection{The entire distribution}
	The number of FRBs in the local universe per unit volume, per unit time, with isotropic specific-energy\footnote{Specific-energy refers to energy per unit frequency.}, $E_{\nu_0}$, at frequency $\nu_0$ is found to be a power-law function, e.g. \cite{LuPiro19}
	\begin{equation}
	f(>E_{\nu_0}) = \phi_{\rm FRB}\, E_{{\nu_0},32}^{-\alpha_{\rm E}}, 
	\label{frb_dens_0}
	\end{equation}
	where $\phi_{\rm FRB} \sim 10^{2.6}$ Gpc$^{-3}$ yr$^{-1}$, $\alpha_{\rm E}=0.7$ and $E_{\nu_0,32}$
	is the isotropic equivalent specific-energy release by bursts at frequency $\nu_0 = 1$ GHz in units of 10$^{32}$ erg Hz$^{-1}$. This power-law function 
	is taken to hold above a minimum FRB energy $E_{\nu_0}^{\rm min}\sim 10^{30}$ erg Hz$^{-1}$ and below $E_{\nu_0}^{\rm max}\sim 10^{34}$ erg Hz$^{-1}$ \footnote{Note that we make no assumption regarding whether the distribution of FRB energies extends to values lower than $E_{\nu_0}^{\rm min}$, as indeed suggested by FRB 200428 \citep{LKZ2020}. Such bursts will only be detectable from $z\gtrsim 1$ even with the more optimistic of the fluence thresholds we adopt below. They will therefore have a negligible effect on the total number of detected FRBs (changing only slightly the low DM side of the distribution) and no affect on the shape of the $dN/d{\rm DM}$ distribution at $\rm{DM}\gtrsim 5000\mbox{pc cm}^{-3}$, relevant for probing the H-reionization era. Similarly, we find that even with our more conservative detection criterion, FRBs at $z\lesssim 10$ can be detected if they have $E_{\nu_0}\gtrsim 1.4 \times 10^{33}\mbox{erg Hz}^{-1}$, well below the value of $E_{\nu_0}^{\rm max}$ adopted above. Since $f(>E_{\nu_0})$ decreases with $E_{\nu_0}$, the implication is that we will be very minimally affected by the uncertainty in the specific location of the maximum cutoff, discussed in \citealt{Luo2020,Wadiasingh2020}.}
	We have assumed here that the spectral energy distribution of FRBs is independent of redshift, which is consistent with current observations \citep{Hashimoto2020}. This assumption can be easily relaxed if future observations, with a much larger sample of FRBs (particularly those with known redshifts), suggest redshift evolution of the FRB luminosity function. 
	
	We take the FRB rate per unit comoving-volume at redshift $z$ as 
	\begin{equation}
	\dot{n}_{\rm FRB}(z, >E_\nu) =  f(>E_\nu) {\dot{n}_*(z)\over \dot{n}_*(z=0)}.
	\label{frb_dens_1}
	\end{equation}
	where $\dot{n}_*(z)$ is the number of stars formed per year at $z$ with mass in the appropriate range so that their remnants are neutron stars; we assume that the initial mass-function (IMF) is the same at low and high redshifts. The total mass of stars formed per comoving-volume per year is taken to be as given by \cite{MD2014} (this assumption is relaxed in \S \ref{sec:MonteCarlo} where we use a star formation rate as a function of galaxy stellar mass and redshift as constrained by observations)
	\begin{equation}
	\label{eq:SFR}
	\dot{m}_*(z) = 0.015 {(1+z)^{2.7}\over 1 + [(1+z)/2.9]^{5.6}} \,
	{\rm M_\odot}\, {\rm year}^{-1} \, {\rm Mpc}^{-3}.
	\end{equation}
	For a non-evolving IMF, 
	\begin{equation}
	{\dot{n}_*(z)\over \dot{n}_*(z=0)} = {\dot{m}_*(z) \over \dot{m}_*(z=0)}.
	\end{equation}
	
	The total number of FRBs per unit time (in observer frame) and per unit DM is
	\begin{equation}
	{d \dot{N}_{\rm FRB} \over d\mbox{DM}} = {\dot{n}_{\rm FRB}(z, \, >E_{\nu})\over
		1 + z(\mbox{DM})}\, 4\pi r^2(\mbox{DM}) {dr\over d\mbox{DM}},
	\label{dN_frb}
	\end{equation}
	where we made use of the comoving volume at redshift $z$,
	\begin{equation}
	dV = 4\pi r^2 dr = {4\pi r^2(z) c \over H(z)}\, dz,
	\end{equation}
	$z(\mbox{DM})$ is given by Eq. \ref{dm-z2}, $r(\mbox{DM})$ is the comoving distance to an FRB at redshift $z$ given by
	\begin{equation}
	r = c\int_0^z {dz\over H} = {c\over H_0 \Omega_{m0}^{1/2}} \int_0^z 
	{1\over \left[ (1+z)^3 + \Omega_{\Lambda0}/\Omega_{m0} \right]^{1/2} },
	\label{r-z}
	\end{equation}
	and the factor $(1+z)$ in the denominator of Eq. \ref{dN_frb} converts the rate from the comoving frame at $z$ to the observer frame. 
	\begin{figure}
		\centering
		\includegraphics[width = 0.4\textwidth]{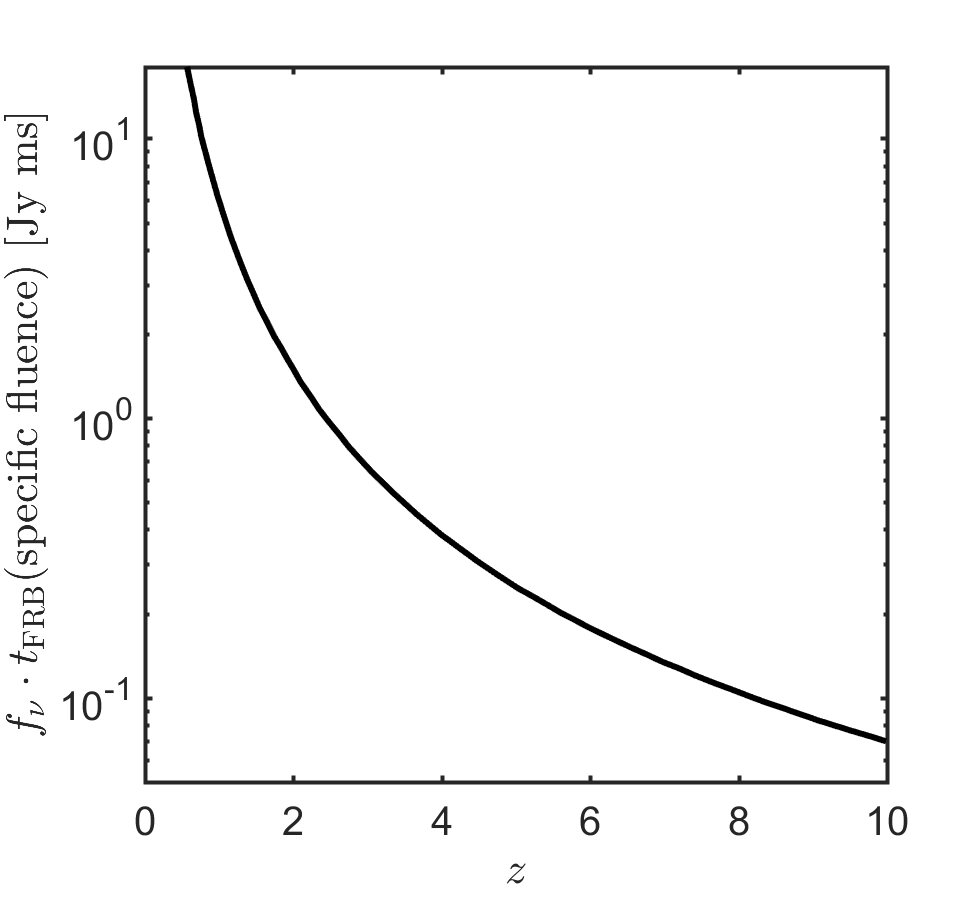}\\
		\includegraphics[width = 0.4\textwidth]{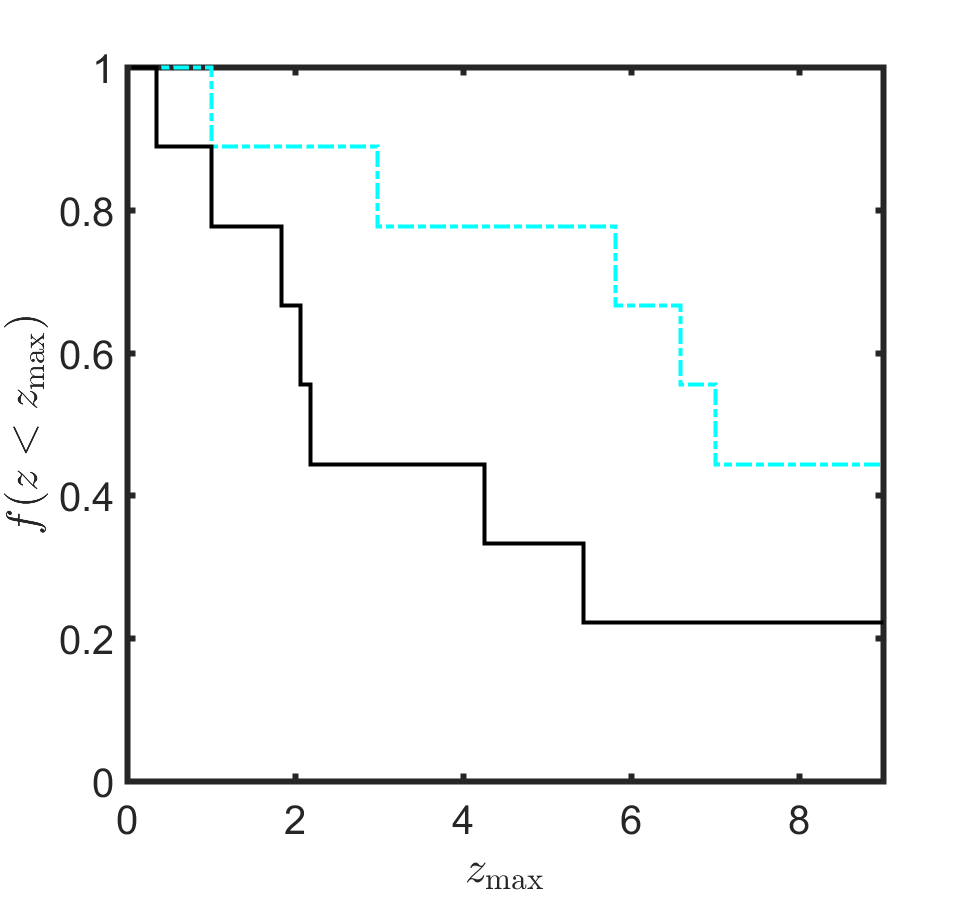}
		\vskip -0.2cm
		\caption{Top: The observed specific fluence ($ f_\nu\times t_{\rm FRB}$) is shown for an FRB at redshift $z$ which has an isotropic equivalent energy per unit frequency of 10$^{32}$ erg Hz$^{-1}$ at 1.0 GHz in its rest frame. The spectrum of the burst is taken to be $L_\nu \propto \nu^{-1.5}$, $t_{\rm FRB}$ is the observed burst duration and the burst is observed at 500 MHz, i.e. $\nu$ in $f_\nu$ is 500 MHz. Bottom: The fraction of 9 FRBs with known redshifts which would be detectable up to a redshift $z$. Results are shown in a solid (dot-dashed) curve for a specific fluence threshold of $e_{\nu}^{\rm o,th}=1$\,Jy ms ($e_{\nu}^{\rm o,th}=0.1$\,Jy ms) at 500 MHz and assuming a spectral slope of $\alpha=-1.5$.
		}
		\label{fig:fluence-z}
	\end{figure}
	
	\subsubsection{The observable distribution}
	\label{sec:obsdistana}
	The DM-distribution of the FRB-rate above the observed specific fluence $e_{\nu}^{\rm o,th}$ is given by
	\begin{equation}
	{d \dot{N}_{\rm FRB}(>e_{\nu}^{\rm o,th}) \over d\mbox{DM}} = {\dot{n}_{\rm FRB}(z, >E_{\nu_1}^{\rm TH})
		\over 1+z(\mbox{DM})} \, 4\pi r^2(\mbox{DM}) {dr\over d(\mbox{DM})},
	\end{equation}
	where 
	\begin{equation}
	\label{eq:fluence}
	E_{\nu_1}^{\rm TH} = 4\pi e_{\nu}^{\rm o,th} r^2, \quad E_{\nu_0}^{\rm TH} = E_{\nu_1}^{\rm TH} 
	(\nu_1/\nu_0)^{\alpha}, \quad \nu_1 = (1+z)\nu 
	\end{equation}
	$\nu_1$ is the frequency in the burst comoving frame that corresponds to the observing band frequency $\nu$, as before $\nu_0 = 1$ GHz (but in the burst frame, at $z$), and $\alpha\sim -1.5$ is the average spectral index for the FRB population. The spectral index of FRBs is still not well determined at this point, due to the relatively narrow band over which most FRBs are observed. Nonetheless, our choice of $\alpha=-1.5$ is consistent with the results of \cite{Macquart2019} for a sample of 23 bright FRBs. Furthermore, \cite{Gajjar2018} have observed bursts from the repeating FRB, 121102, at 4-8 GHz. Their observations demonstrate two important points. First, the intrinsic FRB spectrum can extend up to 8 GHz, which is more than sufficient for the purpose of detecting FRBs from the H-reionization era (i.e. $z\gtrsim 6$) at 0.5 GHz, as assumed for this work. Second, although they do not have simultaneous observations across a wide range of frequencies, \cite{Gajjar2018} find that the maximum flux at $4-8$\,GHz for bursts of FRB 121102 is comparable to the maximum flux observed for bursts at $\sim 1$\,GHz of the same source. This is far from being a conclusive determination of the spectrum, but it suggests that our canonical choice of $\alpha=-1.5$ is likely conservative.

	The observed specific-fluence of an FRB as a function of redshift, for fixed burst parameters in the comoving frame, is shown in the top panel of Fig. \ref{fig:fluence-z}. By substituting the above expression for $E_{\nu_0}^{\rm TH}$ into Eqs. \ref{frb_dens_0} \& \ref{frb_dens_1} we obtain the rate of FRBs per unit volume with intrinsic specific energy at comoving frequency $\nu_1$ that is high enough so that the observed specific fluence is $>e_{\nu}^{\rm o,th}$ at $\nu$. As a proof of the validity of the concept suggested in this paper, we have examined whether any of the FRBs detected so far, would have been detectable if they had originated from the epoch of H reionization (i.e. $z\gtrsim 6$). Such an exercise requires knowledge of the FRB redshift, in order to extrapolate its properties to higher $z$. We consider all the nine FRBs with known redshifts at the time of completion of this work: FRBs 121102, 180916, 180924, 181112, 190102, 190523, 190608, 190611 \& 190711 \citep{Bannister+19,Prochaska+19,Ravi+19,Macquart20}. The fraction of bursts in this sample that would be detectable if they had originated from a higher redshift $z$ is shown in the bottom panel of Fig. \ref{fig:fluence-z}. Taking $e_{\nu}^{\rm o,th}=1\,\mbox{Jy ms}$ ($e_{\nu}^{\rm o,th}=0.1\,\mbox{Jy ms}$) at 0.5 GHz, and $\alpha=-1.5$, we find that 2 (4) out of 9 FRBs with known redshifts would have been detectable at $z\gtrsim 6$. This finding supports the notion that a significant fraction of FRBs during the H reionization epoch, with properties similar to $z \lae 1$ bursts,  should be detectable by current generation of FRB telescopes such as CHIME and ASKAP.

	In our study, we implicitly assume that the criteria for detection is given by a limiting specific fluence. In practice, the threshold for detectability can be more complex. In particular, we do not explicitly account for the smearing of high DM signals. The latter is characterized by a timescale $t_{\rm DM}=8.3\times 10^{-3}\mbox{DM}\,\Delta \nu_{\rm MHz}\,\nu_{\rm GHz}^{-3}$\,ms \citep{Connor2020}, where the DM is measured in $\mbox{pc cm}^{-3}$, $\Delta \nu_{\rm MHz}$ is the spectral width of the channel in MHz and $\nu_{\rm GHz}$ is its central frequency in GHz. The latter two parameters depend on the telescope. For example, for CHIME, $\Delta \nu_{\rm MHz}=0.024, \nu_{\rm GHz}=0.6$ \citep{Connor2020}. If the DM of a burst is large, $t_{\rm DM}$ may become large compared to the intrinsic duration of the burst and the sampling time of the detecting instrument. This will cause a smearing of the burst signal over a time longer than its intrinsic duration. Although the burst fluence is unchanged by this pulse broadening (and by our criteria mentioned above the burst detectability too is unaffected), the background noise fluence increases over this longer duration, and as a result the signal to noise ratio will decrease. At most (when $t_{\rm DM}$ is much larger than the other timescales, which likely only happens when DM is a few thousands or more) the reduction is proportional to $DM^{-1/2}$. Whether this effect is important will depend on the observing telescope and on the intrinsic duration distribution of high-$z$ FRBs. Nonetheless, the discussion above suggests that this might have some effect on the observed distribution.
	
\begin{figure}
		\centering
		\includegraphics[width = .4\textwidth]{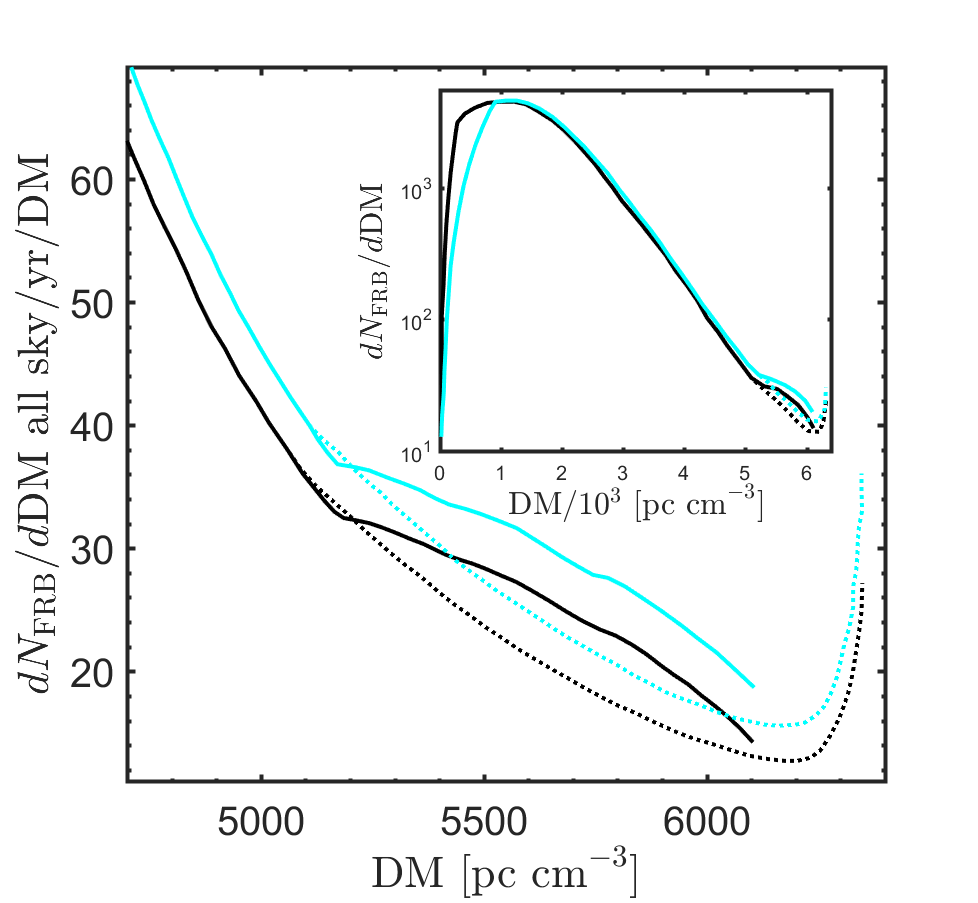}\\
		\includegraphics[width = .4\textwidth]{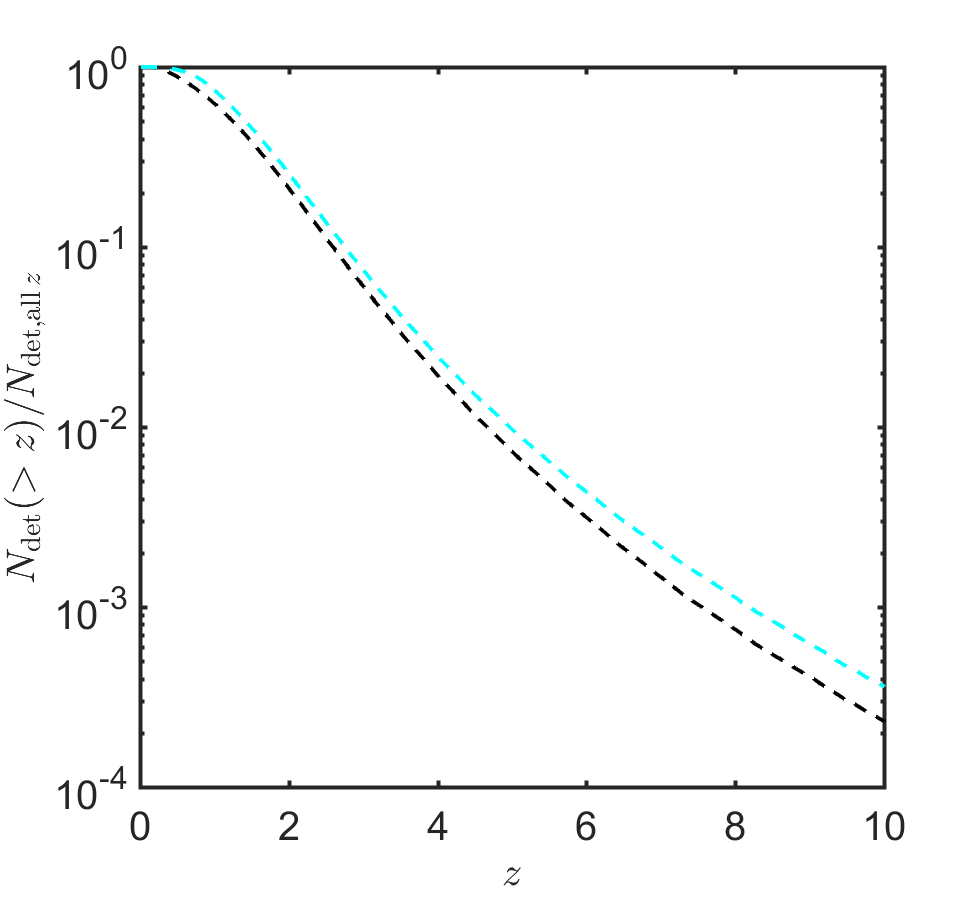}
		\vskip 0.1cm
		\caption{The upper panel shows the number of detected FRBs per unit DM, over the entire sky, per year -- $d \dot{N}_{\rm FRB}/\d\mbox{DM}$ -- for the two different reionization histories described in Fig. \ref{fig:dm-z} (solid lines for $\xi_{\rm e,o}$ and dotted lines for $\xi_{\rm e,t}$), and two different fluence threshold for observations. Black curves show FRBs with observed fluence per unit frequency larger than 1.0 Jy ms at 0.5 GHz, and cyan curves show $0.2 ~{\rm x}~ d \dot{N}_{\rm FRB}/\d\mbox{DM}$ for bursts with observed fluence threshold of 0.1 Jy ms at 0.5 GHz; we have assumed for these calculations that the average FRB spectrum is $f_\nu\! \propto \!\nu^{-1.5}$. The reionization epoch ends at $\mbox{DM}\sim 5{\rm x}10^3\mbox{pc cm}^{-3}$ (see Fig. \ref{fig:dm-z}). $d \dot{N}_{\rm FRB}/\d\mbox{DM}$ is radically different for $\mbox{DM} > 5{\rm x}10^3$ cm$^{-3}$ pc for the two different reionization histories of the universe. The decline of $d \dot{N}_{\rm FRB}/\d\mbox{DM}$ with $\mbox{DM}$ flattens, and could even turn over, during the reionization epoch.  Furthermore, the largest value of $\mbox{DM}$ for FRBs depends on the reionization history as shown in Fig. \ref{fig:dm-z}. The inset in the upper panel provides a zoom-out view of the DM-distribution of the FRB rate between $z$ of 0 \& 10 and as in the main panel cyan curves have been scaled down by a factor 5. The lower panel shows the fraction of detected FRBs that arrive from a redshift $>z$. Results for a fluence threshold of 1 Jy ms (0.1 Jy ms) at 0.5 GHz are shown in black (cyan).
		}
		\label{fig:dN_dDM}
	\end{figure}

	We show in Fig. \ref{fig:dN_dDM} the number of FRBs per unit DM with specific fluence at Earth larger than 1 Jy ms, i.e. $e_{\nu}^{\rm o,th} = 10^{-26}\mbox{erg Hz}^{-1}\mbox{ cm}^{-2}$, at 0.5 GHz; we show $d \dot{N}_{\rm frb}/d\mbox{DM}$ for two cases of hydrogen reionization histories which are the same ones as presented in Fig. \ref{fig:dm-z}. 
	Also shown in Fig. \ref{fig:dN_dDM} is the fraction of detected FRBs in our model (given our limiting fluence criteria) that arrive from a redshift $z$ or greater. For $e_{\nu}^{\rm o,th}=1\,\mbox{Jy ms}$ ($e_{\nu}^{\rm o,th}=0.1\,\mbox{Jy ms}$) we find that $\sim 0.3\%$ ($\sim 0.4\%$) of detected FRBs arrive from $z>6$.
	The FRB rate can also be compared to the magnetar formation rate.
	The neutron star birth rate is calculated from the following IMF \citep{Kroupa2001}
	\begin{equation}
	n_*(m) = A_* \left\{
	\begin{array}{l}
	\hskip -5pt 10.8\, (m/0.08 M_\odot)^{-0.3}    \hskip 36pt m< 0.08\,
	M_\odot \\ \\
	\hskip -5pt (m/0.5 M_\odot)^{-1.3} \hskip 20pt 0.08 M_\odot< m < 0.5\,
	M_\odot \\ \\
	\hskip -5pt (m/0.5 M_\odot)^{-2.3} \hskip 64pt m > 0.5\,
	M_\odot
	\end{array}
	\right.
	\label{kroupa}
	\end{equation}
	Stars leave behind a neutron star remnant if their initial mass lies between $M_1 \approx 8 M_\odot$ and $M_2 \approx 20 M_\odot$. Thus, the birth rate of neutron stars per unit redshift, for the given star formation rate per unit comoving volume of $\dot{m}_*$, is
	\begin{equation}
	{d \dot{N}_{\rm NS}\over dz} = 0.14 \left[ {\dot{m}_*\over  M_\odot} \right]
	\left[ {M_1 \over M_\odot} \right]^{-1.3} \left( 1 - \left[{M_1\over M_2}
	\right]^{1.3}\right) {4\pi r^2\over 1+z} {dr\over dz}.
	\end{equation}
	As before, the factor $(1+z)$ in the denominator converts the comoving frame rate at $z$ to the observer frame. 
	The NS birth rate is similar for \cite{MillerScalo79} and \cite{Chabrier2003} IMFs as they have identical shapes for $m > 1 M_\odot$ and the average stellar mass for both these distributions are about $1 M_\odot$. The fraction of neutron stars that are born with magnetar strength magnetic field, i.e. surface field stronger than 10$^{14}$G, is estimated to be $\sim 40$\% in our galaxy \citep{Beniamini2019}.
	We assume that this fraction at $z>6$ is similar. Therefore, the magnetar birth rate is $\sim 0.4 \, d\dot{N}_{\rm NS}/dz$.
	The FRB rate used in this work, taken from \cite{LuPiro19}, turns out to be such that one in about ten magnetars produces a single burst with specific-energy larger than 10$^{30}$ erg Hz$^{-1}$ in their entire active lifetime. Thus, the FRB rate we have used in this work is highly conservative.
	
	We point out a couple of the most prominent features in Fig. \ref{fig:dN_dDM}. $d \dot{N}_{\rm FRB}/\d\mbox{DM}$ declines for $\mbox{DM} \gae 1.5{\rm x}10^3$ during the era when hydrogen is almost completely ionized and when the star formation rate is decreasing with redshift. The decline stops during the reionization epoch, $\mbox{DM}\gae 5{\rm x}10^3\mbox{ pc cm}^{-3}$, and could even turn into a rising curve depending on the reionization history. This change in the shape of $d \dot{N}_{\rm FRB}/\d\mbox{DM}$ should be possible to detect in the observed FRB DM distribution, and it carries information regarding how reionization proceeded with redshift. The total number of FRBs during the reionization epoch per year, over the entire sky, with observed specific-fluence larger than 1 Jy ms at 1 GHz, is of order 10$^4$ (Fig. \ref{fig:dN_dDM}). So an observing plan that covers a few percent of the sky with threshold specific-fluence of 1 Jy ms should be able to detect of order 10$^3$ FRBs in a few years from the reionization epoch and provide a high signal for exploring how the reionization progressed with redshift. There are several elements in our analysis that have a fair degree of uncertainty (such as fraction of neutron stars that form magnetars or the rate of FRB production at a given energy). However, these uncertainties mainly affect the normalization of the $dN_{\rm FRB}/d\mbox{DM}$ curves, and not so much the functional form of $dN_{\rm FRB}/d\mbox{DM}(z)$ which is what we use to constrain the reionization of the universe.
	
	It has been pointed out by \cite{Zhang2018} that Five hundred meter Aperture Spherical Telescope (FAST) -- which has a threshold fluence sensitivity of 10 mJy ms -- can detect FRBs with isotropic specific-energy of $\sim 10^{32}$ erg Hz$^{-1}$ at $z\sim 10$.
	The estimates we have provided here are
	for a much smaller size telescope (array) with threshold sensitivity of $\sim$1 Jy ms, which should be able to detect brighter FRBs at $z\sim10$.

	If redshift could be measured for a sub-sample of FRBs, then we can determine the energy distribution function $f(>E_{\nu})$ and find out whether it evolves with $z$ or not. This evolution can then be used self-consistently for a larger sample of bursts to calculate $d \dot{N}_{\rm FRB}/d\mbox{DM}$, and that should improve the accuracy of mapping the reionization epoch.
	
	For the simplified model presented in this section, we have ignored the contributions to the DM from the FRB host galaxy ($\mbox{DM}_{\rm gal}$) and the circum-galactic medium ($\mbox{DM}_{\rm CGM}$). We also did not include the density fluctuations in the IGM. We show in \S\ref{fire-DM} that the sum of these contributions to the DM are less than about $1000\mbox{ pc cm}^{-3}$ in the rest frame of the burst.
	For a FRB at redshift $z$, these DM contributions are suppressed by a factor $(1+z)$. Hence, for bursts from the H reionization era, $\mbox{DM}_{\rm gal} + \mbox{DM}_{\rm CGM} \lae 150$ cm$^{-3}$ pc in the observer frame. Thus, these uncertain contributions to the DM during the hydrogen-reionization epoch are about 3\% of the total observed DM for the burst, and the simplified analytical calculations described in this section regarding how FRBs can probe the reionization epoch should be secure.
	
	In \S \ref{sec:MonteCarlo} we provide a more accurate calculation of the DM-distribution of FRBs that makes use of FIRE simulation results. These simulations provide the host galaxy and CGM contributions to the DM, and the star formation rates as a function of galaxy mass \& redshift. Moreover, we also include the fluctuations to the DM$_{\rm IGM}$ from large scale cosmological simulations. Before turning to this analysis we present a second useful probe of H reionization.

	\subsection{$\mbox{DM}_{\rm max}$: a probe of the reionization epoch}
	\label{sec:dmmax}
	
\begin{figure}
		\centering
		\includegraphics[width = .45\textwidth]{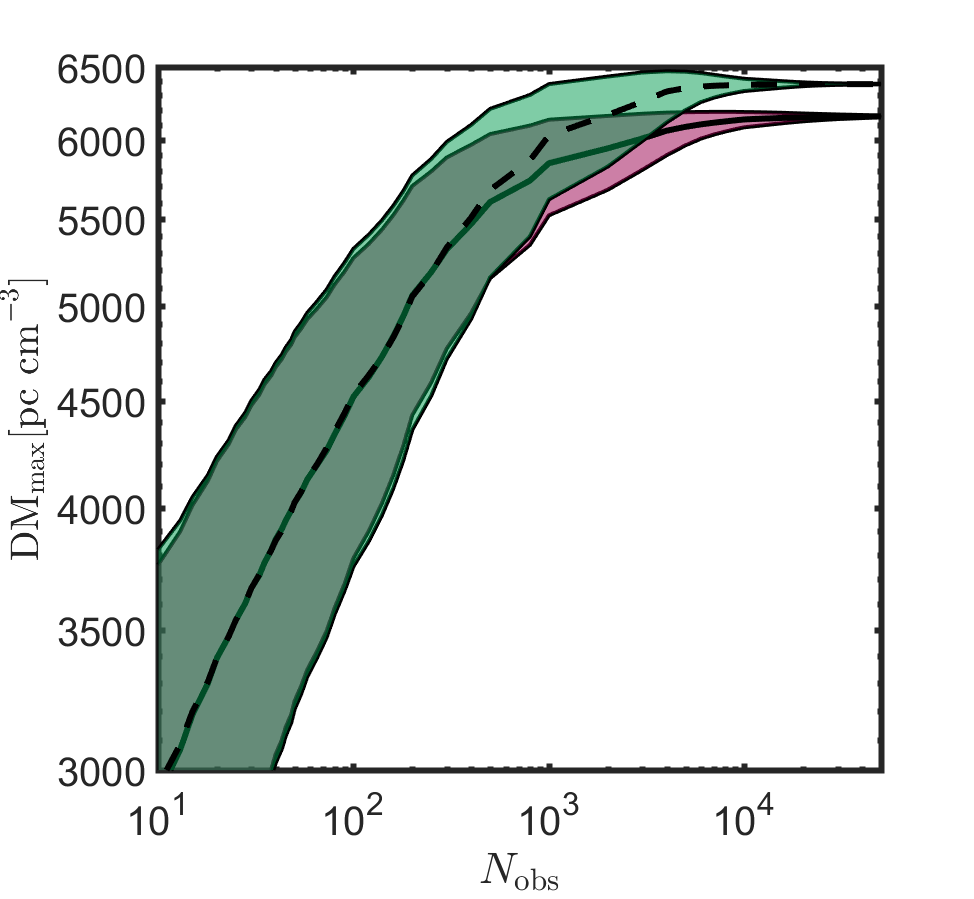}
		\caption{The maximum value of DM, $\mbox{DM}_{\rm max}$, due to electrons in the IGM (assuming no fluctuation in the IGM density), as a function of the number of observed FRBs, $N_{\rm obs}$. The shaded region represent $1\sigma$ statistical scatter about the median value. The purple curve centered on the solid line is for the ionization fraction given by $\xi_{\rm e,o}(z)$ \protect\citep{Robertson2015}, while the green curve centered on the dashed line is for $\xi_{\rm e,t}(z)$ (Eq. \ref{xi-test}).}
		\label{fig:DMmax}
	\end{figure}

	The smallest redshift ($z_{\rm min}$) when hydrogen was almost completely neutral sets the maximum value of DM, i.e. $\mbox{DM}_{\rm max}$, (Figs. \ref{fig:dm-z} \& \ref{fig:dN_dDM}). This is a property that may be relatively easily measured by observers. It translates to a measurement of $z_{\rm min}$ and can be used to constrain $\xi_e(z)$. This approach of using $\mbox{DM}_{\rm max}$ to investigate reionization history is independent of any assumptions regarding the uncertain FRB luminosity function and its potential evolution with redshift (provided that there are FRBs at $z>6$). As a demonstration, we show in Fig. \ref{fig:DMmax} the value of $\mbox{DM}_{\rm max}$ (including statistical scatter, and accounting for the IGM propagation but for clarity taking no error associated with density fluctuations in the IGM) as a function of the number of observed FRBs. Under these assumptions, $\sim5\times 10^3$ FRBs need to be detected to be able to discriminate between the two reionization histories presented in Fig. \ref{fig:dm-z} on the basis of $\mbox{DM}_{\rm max}$. As we will show in the next section, realistic accounts of the errors in $\mbox{DM}_{\rm IGM}$ and the host galaxies' contribution, make the separation between reionization histories more difficult. Nonetheless, this approach of using $\mbox{DM}_{\rm max}$ can be very useful particularly when FRBs with DM$\gae$5x$10^3\mbox{ pc cm}^{-3}$ can be followed up optically to spot their host galaxies and remove those bursts from the sample which are nearby ($z\lae 3$). This will significantly reduce the uncertainty involved in the $\mbox{DM}_{\rm max}$ technique (recall that the ISM and CGM contributions of high $z$ galaxies are suppressed by $1+z$ in the observer frame).
	
\begin{figure}
		\centering
		\includegraphics[width = .45\textwidth]{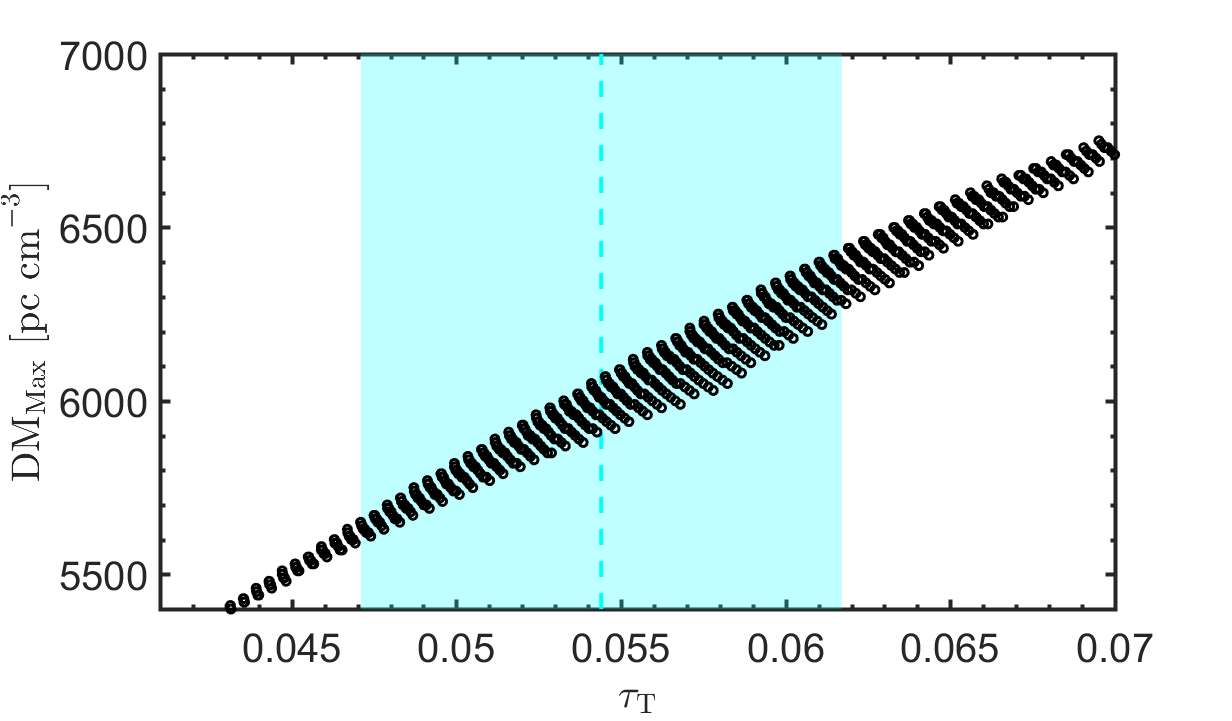}\\
		\includegraphics[width = .45\textwidth]{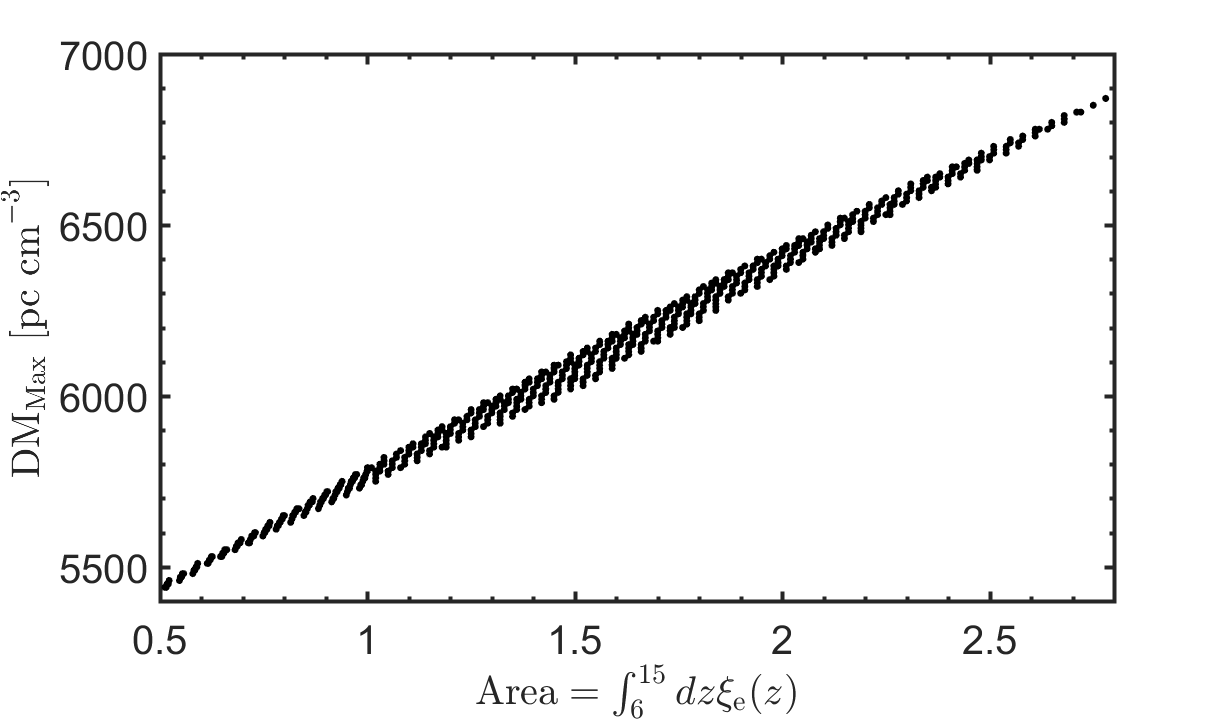}\\
		\includegraphics[width = .45\textwidth]{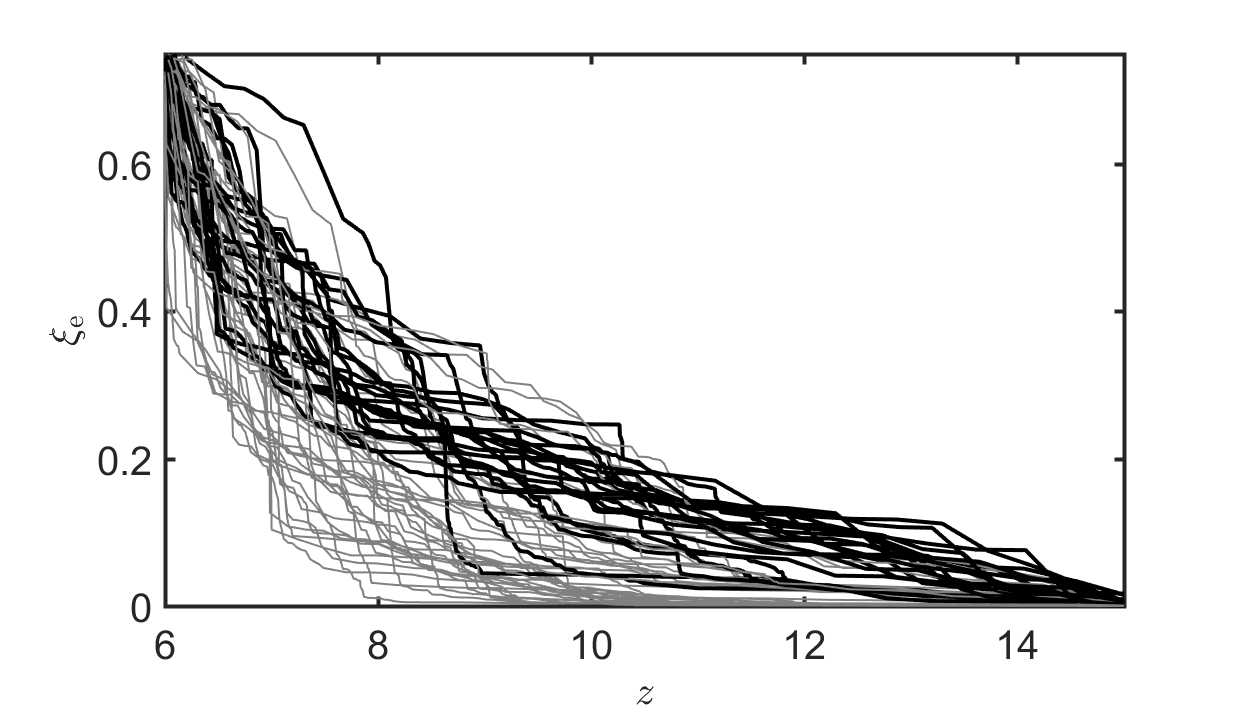}
		\caption{$\mbox{DM}_{\rm max}$ (IGM contribution) vs. 
			Thomson optical depth between us and the last scattering surface, $\tau_{\rm T}$, for different H-reionization histories (top panel). The correlation between the two suggests that measurement of $\mbox{DM}_{\rm max}$ can be used to constraint the reionization history, similar to what has previously been done using $\tau_{\rm T}$. Reionization histories
			that give $\tau_{\rm T} = 0.0544 \pm 0.0073$, consistent with Planck measurement \citep{Planck2020} are marked with a cyan shaded region. This panel also shows that  if the error in the measurement of ${\rm DM_{\rm max}}$ is $\lae 500$ pc cm$^{-3}$ then that would provide a more accurate measurement of $\tau_{\rm T}$ than Planck 2020. The middle panel quantifies the relation between $DM_{\rm max}$ and reionization history; the X-axis is $\int_6^{15} \xi_{\rm e}(z)dz $, i.e. the total area of electron-per-baryon for $6 < z < 15$. The bottom panel shows examples for $\xi_{\rm e}(z)$ curves (in gray) that are consistent with the Planck 2020 measurement of $\tau_{\rm T}$. Thick black curves are consistent with $\tau_T$ (Planck 2020 data) and with $\mbox{DM}_{\rm max}=6000\pm100\mbox{pc cm}^{-3}$.
		}
		\label{fig:DMmaxtauT}
	\end{figure}

	An accurate measurement of $\mbox{DM}_{\rm max}$ can improve the constraint on the reionization provided by Planck satellite's determination of Thomson optical depth to the last scattering surface for CMB photons ($\tau_{\rm T}$). Both quantities are integrals over $n_{\rm e}(z)/H(z)$ with different weights of ($1+z$); compare Eq. \ref{dm-z1} to the equation below for $\tau_{\rm T}$
	\begin{equation}
	\label{eq:tauT}
	\tau_{\rm T}(z) = \sigma_T c\int_0^z dz' {n_{\rm e}(z') \over (1+z') H(z')}.
	\end{equation}
	As a result, the two quantities are correlated for different reionization histories. Fig. \ref{fig:DMmaxtauT} explicitly shows this correlation for a complete ensemble of $\xi_e(z)$ between redshifts of 6 \& 15. The value of $\tau_{\rm T}$ measured by the Planck satellite observations is $0.0544\pm 0.0073$ \citep{Planck2020}. Fig. \ref{fig:DMmaxtauT} shows that if 
	$\mbox{DM}_{\rm max}$ could be measured with an accuracy that is better than $\sim500~ \mbox{pc cm}^{-3}$ then that would improve the constraint on the reionization history provided by Planck. According to Fig. \ref{fig:DMmax}, this is a very realistic goal for FRB surveys.
	
	\section{FRB DM distribution and reionization: Monte Carlo simulation}
	\label{sec:MonteCarlo}
	To explore the $d \dot{N}_{\rm FRB}/\d\mbox{DM}$ distribution more accurately, we make use of results from the FIRE simulations of high redshift galaxies \citep{Ma2018,Ma2020} and connect them with observational constraints on galaxy properties at low redshift \citep{Behroozi2019}. We also use the results of \cite{Jaroszynski2019}, that has analyzed the {\it Illustris} simulations, to determine the scatter of $\mbox{DM}_{\rm IGM}$. We summarize below the ingredients adopted from these simulations.
	
	\subsection{Contributions to DM from FRB host-galaxy and CGM, and the IGM}
	\label{fire-DM}
	
\begin{figure}
		\centering
		\includegraphics[width =.4\textwidth]{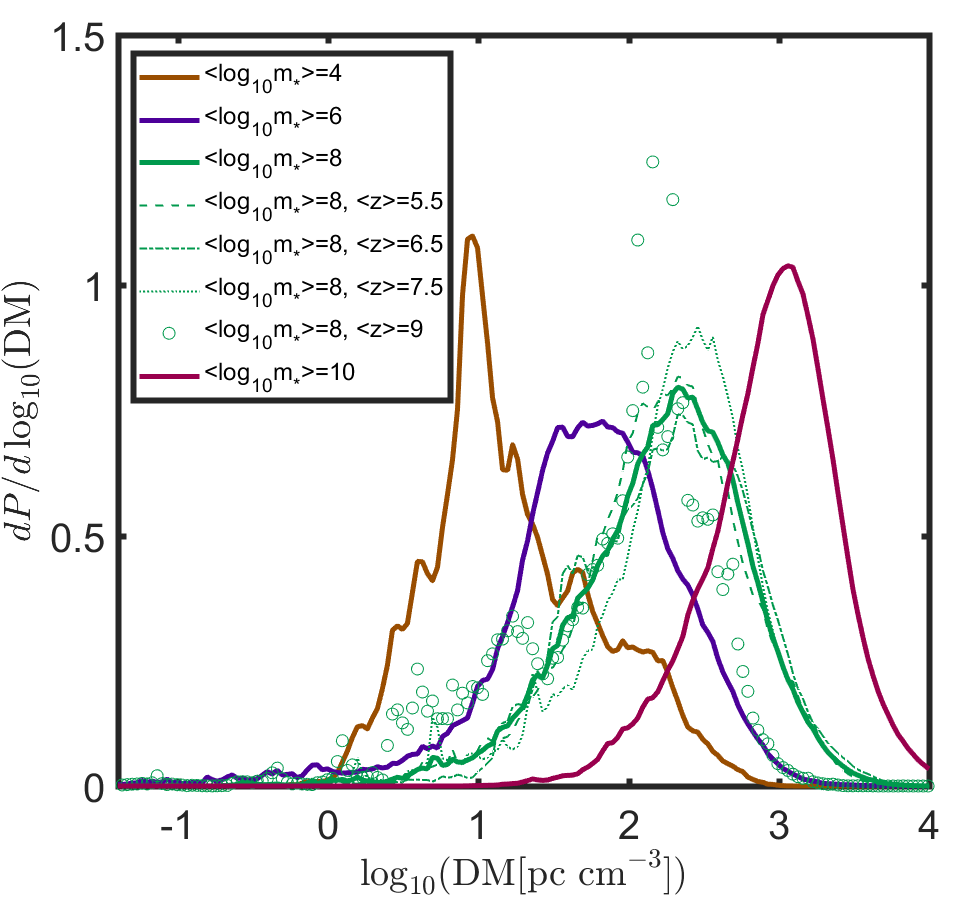}
		\includegraphics[width = .4\textwidth]{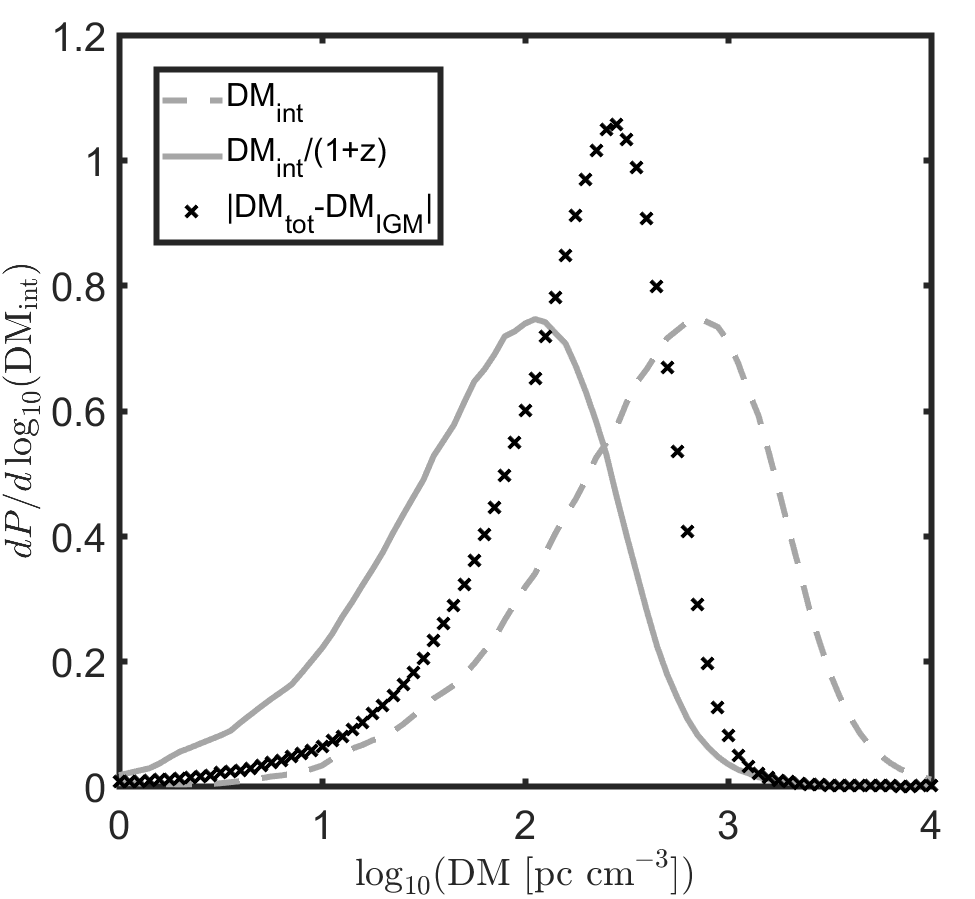}
		\caption{Top: distribution of intrinsic (ISM +CGM) contribution to DM from high redshift galaxies in the FIRE simulations. Thin lines depict distributions dependent on both galaxy stellar mass (within bins of $\Delta \log_{10}(m_{*})\pm 0.5$) and redshift (within bins of $\Delta z\pm 0.5$), while thick lines depict the distributions averaged over redshift (in the range $z=5-12$), but still dependent on galaxy stellar mass. Bottom: A gray line shows the distribution averaged over both galaxy mass and redshift (integrated over the range of $m_*$ and $z$ explored by the FIRE simulations, $\log_{10}(m_*/M_{\odot})=3.5-10.5$ and $z=5-12$), as weighted by the star formation and stellar mass functions accordingly (see \S \ref{sec:weightedDMdist} for details). We present both the distribution in the galaxy comoving frame (dot-dashed) and the distribution in the observed frame, in which the DM from a galaxy at $z$ is reduced by a factor of $1+z$ (solid). The overall distribution of the error in $\mbox{DM}_{\rm IGM}(z)$, given by $|\mbox{DM}_{\rm tot}-\mbox{DM}_{\rm IGM}|$ weighted by the star formation and galaxy mass functions and integrated over all redshifts and all stellar masses is depicted by black Xs. We stress that these `averaged' distributions shown in the bottom panel are presented for clarity and are not directly used in our Monte Carlo calculation presented in \S \ref{sec:MonteCarlospecific}.}
		\label{fig:dPdDM}
	\end{figure}

	The FIRE simulations provide us with estimates for the electron column densities in the ISM and CGM of high redshift ($z>5$) galaxies (we will denote the sum of these two components as $\mbox{DM}_{\rm int}\equiv \mbox{DM}_{\rm gal}+\mbox{DM}_{\rm CGM}$).
	We have extracted this data from the simulations in \cite{Ma2020}. The simulated sample consists of $\sim 8500$ galaxies in $m_{\ast} \sim 10^{3.5}$--$10^{10.5}\,M_{\sun}$ at $z\sim$5--10, with gas ionization states determined by Monte Carlo radiative transfer calculations (section 2.3 therein). In this work, we assume only star particles between 3 and 30 Myrs old produce FRBs (at a constant rate) and calculate the HII column densities from these particles out to the halo virial radius along 10 random sight-lines.
	The results are used to obtain the probability distribution $dP/d\mbox{DM}_{\rm int} (\mbox{DM}_{\rm int}|m_{*},z)\equiv dP_{m_*,z}/d\mbox{DM}_{\rm int}$ where $m_{*}$ is the stellar mass of the galaxy. 
	We directly adopt the numerical probability functions from the simulations
	in our Monte Carlo simulations as depicted in Fig. \ref{fig:dPdDM}. 
	The ISM of high-$z$-galaxies can be highly turbulent, and it is not uncommon for them to have star formation in {\it bursts}. This may lead to dense shells or at the front of {\it superbubbles} formed by supernova shock compression, and DM values as large as $10^4\mbox{pc cm}^{-3}$ in the host galaxy frame (the latter seen for galaxies with $m_*\gtrsim 10^{10}M_{\odot}$). FIRE simulations find the dependence of DM distributions on redshift to be weak for $5<z<10$ (see discussion below and Fig. \ref{fig:dPdDM}). For this reason we use the same distributions for galaxies in the redshift range $z=3-5$. However, at lower redshifts, $z=0-3$, the high DM tail of these distributions discussed above is no longer reliable for several reasons: (a) lower gas fraction in those galaxies, (b) effectively weaker feedback (as star formation occurs in rotating disks and supernova bubbles are confined), (c) a large fraction of the most massive galaxies ($m_*\gtrsim 10^{10}M_{\odot}$) are quenched by $z\sim 0$. We have used the FIRE simulations at $z=0$ to verify that the median DM from galaxies of $m_*\approx 10^{10}M_{\odot}$ is decreased by a factor of $\sim 10$ compared to galaxies with similar stellar mass at $z>5$.  
	Furthermore, at these lower redshifts, we have useful empirical evidence regarding contributions of FRB host galaxies to the DM. As discussed in more detail in \S \ref{sec:immediateenv}, all 9 FRBs with known $z$ -- for which the IGM contribution to the DM is bounded -- have $\widetilde{\mbox{DM}}_{\rm ex}\lesssim 200\,\, \mbox{pc cm}^{-3}$. Finally, at these lower redshifts, the hosts can be identified by observers and any rare case with abnormally large DM values can be removed from the analysis we are suggesting in this work. For all these reasons, for $0<z<3$ we adopt the same DM distributions as at higher $z$, but cutoff the distributions at\footnote{We have also verified that our results are not qualitatively affected if we instead take the cutoff at $\mbox{DM}=1000\,\mbox{pc cm}^{-3}$.} $\mbox{DM}=200\,\mbox{pc cm}^{-3}$.
	
	For clarity, and ease of use, we present here an approximate fitting functions for 
	$dP_{m_*,z}/d\mbox{DM}_{\rm int}$, used for $z>3$, which are reasonably represented by 
	log-normal distributions:
	\begin{equation}
	\frac{dP_{m_*,z}}{d\log_{10}(\mbox{DM}_{\rm int})}=\frac{1}{\sigma\sqrt{2\pi}}\exp\bigg[{-\frac{1}{2}\bigg(\frac{\log_{10}(\mbox{DM}_{\rm int})-\mu_{m}}{\sigma}\bigg)^2}\bigg]
	\end{equation}
	with $\mu_m=0.315\log_{10}(m_{*})-0.221$, and a roughly constant width 
	$\sigma\approx0.51$.
	
	For cosmological studies, $\mbox{DM}_{\rm int}$ is a source of error, that obscures the direct relationship between $\mbox{DM}_{\rm IGM}$ and $z$. An additional error arises due to fluctuations in the electron density in the IGM so that values of $\mbox{DM}_{\rm IGM}$, at a given $z$, along different lines of sights, are different. \cite{Jaroszynski2019} has analyzed electron column density distribution of the {\it Illustris} large scale cosmological simulations, and estimated the fluctuations to the DM-IGM to be given by
	\begin{equation}
	\sigma_{\mbox{DM}_{\rm IGM}}\approx 0.13\, \mbox{DM}_{\rm IGM}(1+z)^{-1/2}
	\label{del-DM-igm}
	\end{equation}
	for $0\lesssim z\lesssim 3$. We note, however, that 
	$\sigma_{\mbox{DM}_{\rm IGM}}$ is highly uncertain\footnote{More recently, \cite{Jaroszynski2020}, using the results of the same {\it Illustris} simulations find that in the range of redshift $0\leq z\leq 3$, a better 
		fit is given by $\sigma_{\mbox{DM}_{\rm IGM}} \approx 0.2\mbox{DM}_{\rm IGM}\, z^{-1/2}$.}, especially at large redshift. For consistency with the published results we adopt $\sigma_{\mbox{DM}_{\rm IGM}}$ as given by 
	Eq. \ref{del-DM-igm} for all $z$. We assume a Gaussian distribution of $\mbox{DM}_{\rm IGM}$ with a median value given by Eq. \ref{dm-z2} and a standard deviation according to $\sigma_{\mbox{DM}_{\rm IGM}}$ above. We note that in the future, when the redshifts of many high $z$ FRBs is measured, $\mbox{DM}_{\rm IGM}$ and $\sigma_{\mbox{DM}_{\rm IGM}}$ would be empirically determined. 
	
	The distribution of $|\mbox{DM}_{\rm tot}-\mbox{DM}_{\rm IGM}|$ (including contributions from $\sigma_{\mbox{DM}_{\rm IGM}}$, which could be positive or negative and from $\mbox{DM}_{\rm int}/(1+z)$ which is positive by construction) is presented in Fig. \ref{fig:dPdDM}, and its median value is found to be $\sim 150\mbox{ pc cm}^{-3}$. This is true for the distribution weighted over redshift according to the SFR. For the distribution within specific redshift bins the median increases only modestly with redshift (reaching a median value of $\sim 300\mbox{ pc cm}^{-3}$ at $z\sim 6$). This suggests that the DM measurements can be used to estimate FRB redshifts with $\sim 10$\% error, for $\mbox{DM} > 3\times10^3\mbox{pc cm}^{-3}$, i.e. for $z$ between 3 and 6. Note that at $z>6$ this technique fails because of the intrinsic uncertainty in $\mbox{DM}_{\rm IGM}$ arising from the reionization history of the universe.
	
	\subsection{Star formation rate dependence on galaxy mass}
	\label{SFRandstellarmassfunc}
	The distribution of star formation rate as a function of galaxy stellar mass, $\rm{SFR}(m_*|z)$, for low redshift galaxies ($z<4$) is taken from \cite{Behroozi2019} (Fig. 3 of that paper) while for high redshift galaxies ($z>5$) it is taken according to the PL fitting function found by \cite{Ma2018} (Eq. 5 of that paper). We adopt a smooth interpolation in the intermediate redshift regime.
	Similarly, for the stellar mass functions, $\Phi(m_*|z)\equiv m_*\frac{dN}{dVdm_*}$, we adopt the relations given by \cite{Behroozi2019} (Fig. 3 of that paper) for low redshift galaxies ($z<4$), and use the results from \cite{Ma2018} (e.g. Fig. 9 of that paper) for high redshift galaxies ($z>5$). A smooth interpolation in the intermediate redshift regime is adopted. 
	In principle, the functions $\rm{SFR}(m_*|z)$ and $\Phi(m_*|z)$ can be used to find the total star formation rate at an arbitrary $z$:
	\begin{equation}
	\label{eq:SFRint}
	\dot{m}_{*}(z)=\int \frac{\Phi(m_*|z)}{m_*} SFR(m_*|z)dm_*.
	\end{equation}
	In practice, it is well established in the literature, e.g. \cite{Behroozi2019}, that this procedure does not yield perfect agreement with the empirically determined star formation history expressed by Eq. \ref{eq:SFR}.
	We make a comparison between the two functions and find they are in agreement to within $\sim 20\%$ in the range $0\leq z\leq 9$.

	
	\subsection{Star formation rate weighted $\mbox{DM}_{\rm GAL}$ \& $\mbox{DM}_{\rm CGM}$: FIRE 
		simulations}
	\label{sec:weightedDMdist}
	For clarity, we provide below an approximate measure of the overall internal DM contributions as weighted by the star formation rate and galaxy mass functions discussed above. We stress that this `averaged' distribution is not directly used in our Monte Carlo calculation described in \S \ref{sec:MonteCarlospecific}, for which we use the full distributions discussed above.
	We combine the functions $dP_{m_*,z}/d\mbox{DM}_{\rm int}$, $\rm{SFR}(m_*|z)$, 
	$\Phi(m_*|z)$ described in \S \ref{fire-DM}, \S \ref{SFRandstellarmassfunc}
	to calculate the overall distribution of $\mbox{DM}_{\rm int,obs}=
	\mbox{DM}_{\rm int}/(1+z)$ integrated over stellar mass and redshift, and
	weighted by the star formation rate and stellar mass functions:
	\begin{eqnarray}
	\frac{dP}{d\mbox{DM}_{\rm int,obs}}\!=\!A\int \!dz \int \!dm_* 
	\frac{\rm{SFR}(m_*|z)}{1+z}\frac{\Phi(m_*|z)}{m_*}\frac{dV}{dz} \frac{dP_{m_*,z}}{d\mbox{DM}_{\rm int,obs}}
	\end{eqnarray}
	where $A$ is a normalization constant set such that the overall probability is unity, the factor of $1\!+\!z$ converts the time measured in the comoving to the observer frame, and $\frac{dP_{m_*,z}}{d\mbox{DM}_{\rm int,obs}}\equiv (1\!+\!z)\restr{\frac{dP_{m_*,z}}{d\mbox{DM}_{\rm int}}}{ \mbox{DM}_{\rm int}(1+z)}$.
	The resulting distribution is shown in the bottom panel of Fig. \ref{fig:dPdDM}.
	
	\subsection{Contribution to DM from immediate FRB environment}
	\label{sec:immediateenv}
	The immediate environment of FRB sources ($10^{10}$--10$^{18}$ cm from the NS), is likely to consist of electron densities much higher than typical ISM values. That being said, the much smaller scale of the former regions results in contributions to the DM that (barring extreme conditions) are expected to be sub-dominant compared to the ISM. For this reason we neglect this contribution in our Monte Carlo simulations discussed below. In this sub-section we demonstrate the above point using both observed FRBs and theoretical considerations.
	
	So far, $\sim 100$ FRBs have been reported in the online FRB catalog\footnote{http://frbcat.org/}. The catalog provides the total DM value observed for these FRBs ($\mbox{DM}_{\rm tot}$) as well as contributions from our Galaxy along the line of sight, $\mbox{DM}_{\rm MW}$, which are calculated using the Galactic electron density model of \cite{CL2002} \footnote{This model accounts for the dispersion measure given by the Milky Way disk, but not the halo. However, as shown by \cite{Yao2017}, the latter only constitutes a minor correction.}. For nine FRBs in the catalog, the redshifts of their host galaxies have been measured. The lowest value of $\mbox{DM}_{\rm tot}$ in the catalog is $103.5\pm 0.7\mbox{pc cm}^{-3}$ for FRB181030, and there are ten additional FRBs with $\mbox{DM}_{\rm tot}<200\,\mbox{pc cm}^{-3}$. These provide the most conservative upper limit on the contributions of the FRB local-environments to the DM, as these values have not been corrected for any of the other contributions to the DM such as from the ISM of the host galaxy or the IGM. Subtracting the contribution to the DM from Milky Way we find, $\mbox{DM}_{\rm ex}=\mbox{DM}_{\rm tot}-\mbox{DM}_{\rm MW} \lesssim 2\,\mbox{pc cm}^{-3}$ for one FRB, and $\mbox{DM}_{\rm ex}<100\,\mbox{pc cm}^{-3}$ for 5 FRBs. 
	
	For bursts with known redshifts, we can subtract the contribution of the IGM to $\mbox{DM}_{\rm tot}$ according to \footnote{The highest measured $z$ for an FRB is $z=0.66$, so there are no uncertainties due to ionized fraction} Eq. \ref{dm-z2} and calculate the excess $\widetilde{\mbox{DM}}_{\rm ex}=\mbox{DM}_{\rm tot}-\mbox{DM}_{\rm IGM}-\mbox{DM}_{\rm MW}$ (see also \citealt{Bhattacharya2020}). We find that for 4 of the 9 bursts with known $z$, $\widetilde{\mbox{DM}}_{\rm ex}\lesssim 60\, \mbox{pc cm}^{-3}$ and for the other 5 $\widetilde{\mbox{DM}}_{\rm ex}\lesssim 200\, \mbox{pc cm}^{-3}$. These excess DM values include both the immediate environment of the FRBs and the contribution from the ISM $\&$ CGM of their host galaxies. These values of $\widetilde{\mbox{DM}}_{\rm ex}$ are in agreement with the contributions from the FRB-host's ISM $\&$ CGM presented in Fig. \ref{fig:dPdDM}. Therefore, observations suggest that the contributions of the immediate environment of FRBs to the DM is $\lesssim 60 \mbox{pc cm}^{-3}$ (and possibly much smaller). 
	
	The recent detection of FRB 200428, from a magnetar in our Galaxy, has clearly demonstrated that at least some FRBs are produced by magnetars. Assuming this to be true for the general FRB population, one can calculate the expected DM contribution of the pulsar wind nebula (PWN) that could surround a young magnetar. The result is (as also calculated by \citealt{Yu2014,YZ2017}, although we note that their calculation was missing a factor $\propto \Gamma(R_{\rm LC})^{-2}$, where $R_{\rm LC}$ is the light-cylinder, arising due to the fact that only a narrow column of the pulsar wind, of thickness $\sim R_{\rm LC}/2\Gamma(R_{\rm LC})^2$ can be surpassed by the radio wave before significantly expanding and diluting)
	\begin{equation}
	\label{eq:DMPWN1}
	\mbox{DM}_{\rm PWN}\!=\!4\times \!10^{-6} \bigg(\frac{\mu_{\pm}}{10^4}
	\bigg)^{{4\over 3}}\!\bigg(\frac{B}{2.2\!\times \!10^{14}\mbox{G}}
	\bigg)^{2\over3}\!\bigg(\frac{P}{3.2\mbox{ s}}\bigg)^{-{7\over3}}
	\mbox{pc cm}^{-3}
	\end{equation}
	where we have scaled the result relative to the observed period, $P$ and the surface magnetic field, $B$, inferred from spin-down for SGR 1935+2154, the Galactic magnetar which was the source of FRB 200428. The pair multiplicity, $\mu_{\pm}$, is not as well determined from observations. Nonetheless, this result demonstrates that the DM contribution from the PWN is highly unlikely to be significant for FRBs. One should keep in mind, however, that FRB 200428 is much less active as compared with the cosmological population of repeaters \citep{MBSM2020}. One possibility is that these more active repeaters arise from much more strongly magnetized, and possibly faster spinning magnetars (see however \citealt{Beniamini+20}). These objects might have a larger $\mbox{DM}_{\rm PWN}$ (see Eq. \ref{eq:DMPWN1}). However, higher values of $B$ and lower values of $P$ lead to faster dipole spin-down times, $\tau_{\rm sd}=3Ic^3P^2 /4\pi^2B^2R^6$ and a fast rotation for such magnetars cannot be maintained for long. The most active repeater, FRB 121102, has been active since its discovery more than 8 years ago. At the very least, the age, $t$, of the FRB source should be larger than this. For $t\gg \tau_{\rm sd}$, one can use the spin evolution to recast Eq. \ref{eq:DMPWN1} as
	\begin{equation}
	\label{eq:DMPWN2}
	\mbox{DM}_{\rm PWN}\!=\!2\times 10^{-3} \bigg(\frac{\mu_{\pm}}{10^4}\bigg)^{4\over3}
	\!\bigg(\frac{B}{10^{15}\mbox{ G}}\bigg)^{-{5\over3}}\!\bigg(\frac{t}
	{10\mbox{ yr}}\bigg)^{-{7\over6}}\mbox{ pc cm}^{-3}
	\end{equation}
	which shows that the PWN of even very young magnetars are expected to make a small contribution to the DM.
	
	The contribution to the DM from the supernova remnant left over from the birth of the magnetar is also small for a system of age larger than a few
	tens of years; it is $\lae 30$ pc cm$^{-3}$ for a 10$^2$ year old remnant with mass 10$^2$ M$_\odot$ if the gas is fully ionized, and much smaller for a remnant which has cooled down due to adiabatic expansion and is largely neutral.
	
	\subsection{Monte Carlo simulations of FRB $\mbox{DM}$ distributions to investigate
		hydrogen reionization epoch}
	\label{sec:MonteCarlospecific}
	Having estimated the expected contributions to the total DM of an FRB 
	from the host galaxy, its circum-galactic medium and the fluctuations in 
	the IGM electron density, we can now extend the calculations presented 
	in \S \ref{sec:FRBdistAna} by taking into account the fluctuating contributions
	from these various components. We do this by means of a Monte Carlo 
	simulation. Our calculation proceeds as follows. We randomly draw an FRB 
	redshift according to the distribution given in Eq. \ref{eq:SFR}. We then 
	randomly assign a stellar mass, $m_*$, to the host galaxy of the FRB using 
	the distribution $\Phi(m_*|z)$ (see \S \ref{SFRandstellarmassfunc}). 
	Next, we obtain the contributions of the host galaxy and its CGM to
	the total DM, i.e. $\mbox{DM}_{\rm int}$, using the distribution function 
	$dP_{m_*,z}/d\mbox{DM}_{\rm int}(\mbox{DM}_{\rm int}|m_{*},z)$ (see \S \ref{fire-DM}). Finally, the spectral energy of the FRB is drawn from the distribution given by Eq. \ref{frb_dens_0}. With these ingredients in place, we can calculate the total $\mbox{DM}$ (accounting for the IGM contribution and the various fluctuations) 
	as detailed in \S \ref{fire-DM}. The fluence of the FRB at the observed 
	frequency is calculated using Eq. \ref{eq:fluence}. This procedure is repeated 
	$N$ times to simulate a population of observed FRBs which lie above the 
	threshold for observations that we have taken as an example to be 
	1 Jy ms at 0.5 GHz. 
	
\begin{figure}
		\centering
		\includegraphics[width = .4\textwidth]{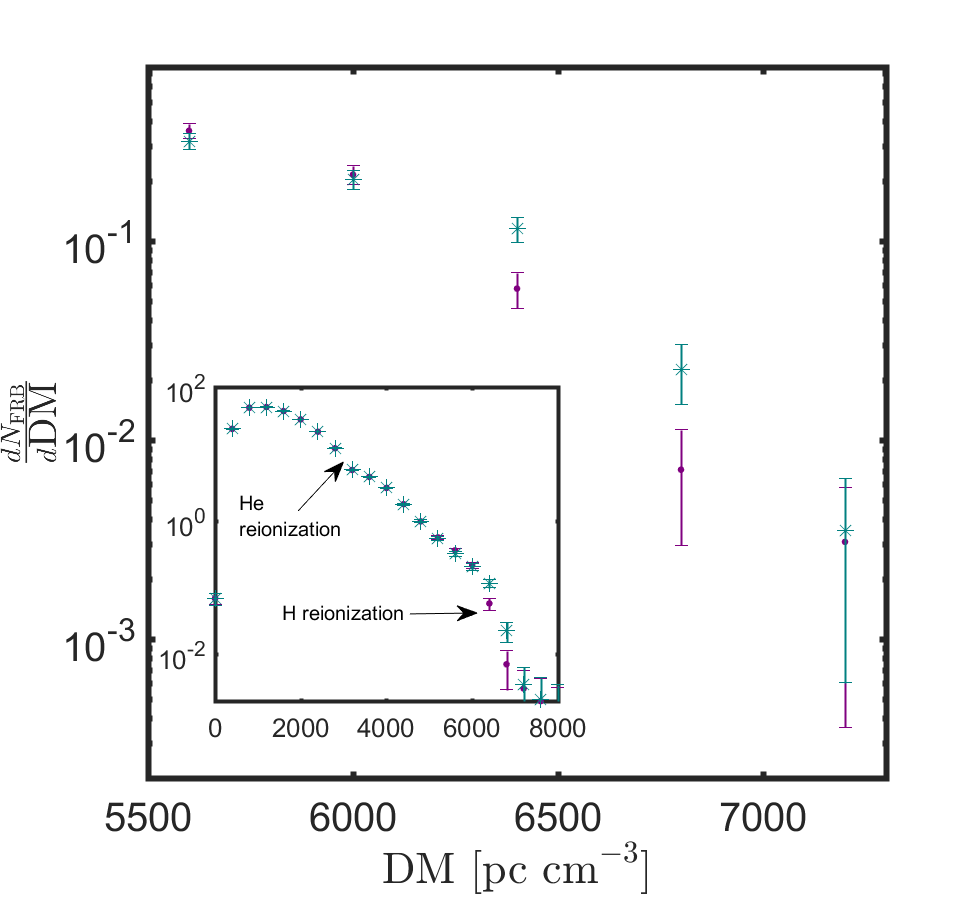}
		\includegraphics[width = .4\textwidth]{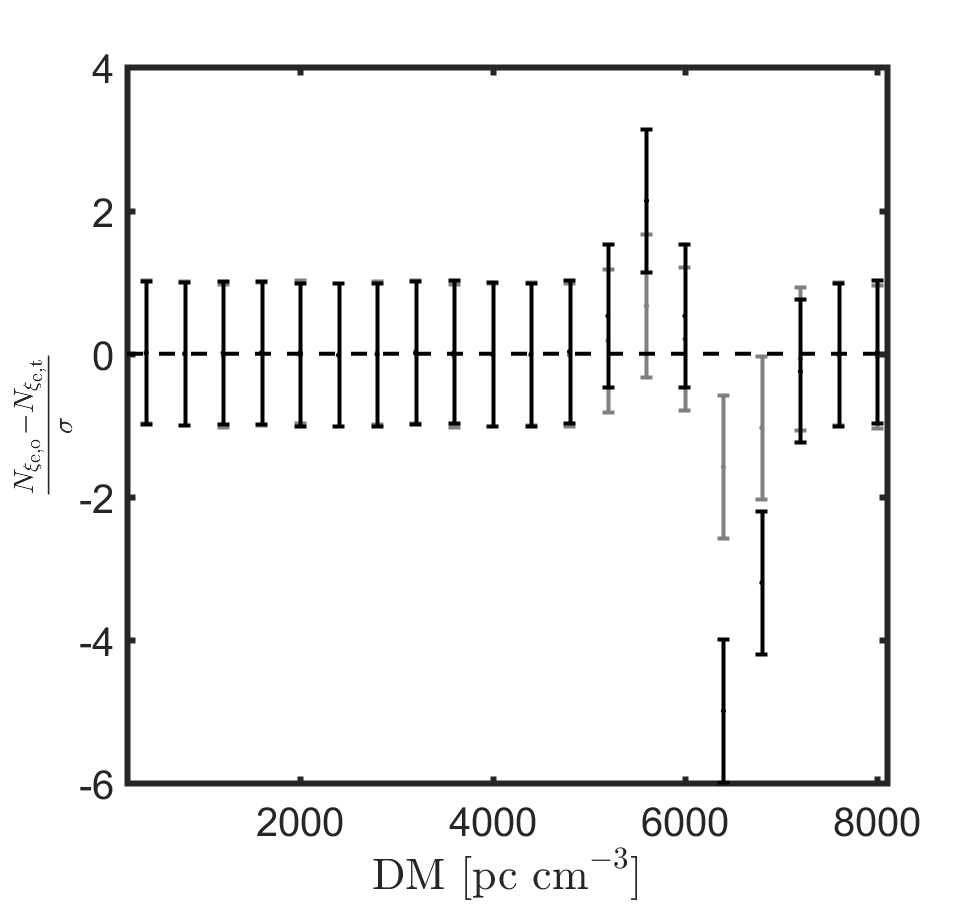}
		\caption{Top: Distribution of $dN_{\rm FRB}/d{\rm DM}$ resulting from a Monte Carlo simulation, with $N=10^5$ detected FRBs, grouped in bins of constant width, $\Delta \mbox{DM}=400\mbox{ pc cm}^{-3}$. The purple (cyan) error bars centered on dots (asterisks) depicts $1\sigma$ fluctuations about the mean value for the ionization fraction as described by $\xi_{\rm e,o}(z)$ \protect\citep{Robertson2015} ($\xi_{\rm e,t}(z)$ defined in \ref{xi-test}). Bottom: Deviations between the model using $\xi_{\rm e,o}(z)$ relative to $\xi_{\rm e,t}(z)$  (Eq. \ref{xi-test}). The former model results in a relative excess of bursts with $5000\lesssim \mbox{DM}\lesssim6000\mbox{ pc cm}^{-3}$ and a dearth of bursts with $6000\lesssim \mbox{DM}\lesssim7000\mbox{ pc cm}^{-3}$. The gray bars depict results with $N=10^4$ detected bursts and the black with $N=10^5$ detected bursts.}
		\label{fig:dNdDMMC}
	\end{figure}

\begin{figure}
		\centering
		\includegraphics[width = .4\textwidth]{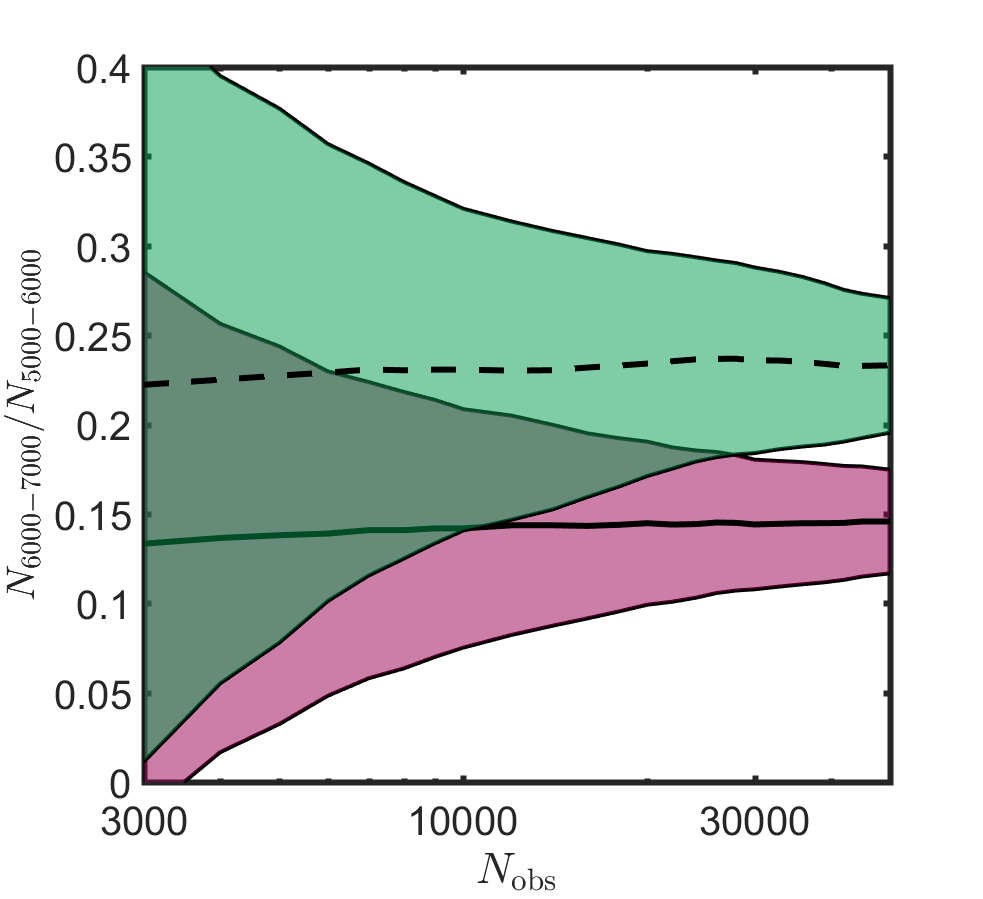}
		\caption{Ratio of detected bursts with $6000\lesssim \mbox{DM}\lesssim7000\mbox{ pc cm}^{-3}$ and those with $5000\lesssim \mbox{DM}\lesssim6000\mbox{ pc cm}^{-3}$, as a function of the total number of detected bursts, $N$, for models with different re-ionization histories (purple region / solid line for $\xi_{\rm e,o}(z)$ \protect\citep{Robertson2015} and green region / dashed line for $\xi_{\rm e,t}(z)$ defined in Eq. \ref{xi-test}).}
		\label{fig:DMbinsratio}
	\end{figure}

	Results of our Monte Carlo simulation are presented in Fig. \ref{fig:dNdDMMC}. The HII re-ionization signature becomes apparent in the DM range $5000-7000\mbox{ pc cm}^{-3}$ (higher DM values are completely dominated by the statistical fluctuations of $\mbox{DM}_{\rm int}$ and $\sigma_{\mbox{DM}_{\rm IGM}}$ for FRBs at $z \gtrsim 3$). With bins of constant width $\Delta \mbox{DM}=400\mbox{ pc cm}^{-3}$, the two models compared in the figure (one is the reionization history as described by $\xi_{\rm e,o}(z)$ \citep{Robertson2015}, and the other one is the test model, $\xi_{\rm e,t}(z)$, given by Eq. \ref{xi-test}) can be distinguished at a level greater than $2\sigma$ with $N\!=\!few\!\times\!10^4$ detected bursts. Since the $\xi_{\rm e,o}(z)$ reionization history model leads to an excess of bursts with $5000\lesssim \mbox{DM}\lesssim6000\mbox{ pc cm}^{-3}$ and a dearth of bursts with $6000\lesssim \mbox{DM}\lesssim7000\mbox{ pc cm}^{-3}$ as compared with the test model, we plot the ratio of bursts in these two DM bins in Fig. \ref{fig:DMbinsratio}. The two re-ionization histories can be differentiated at a $>2\sigma$ level already with a total of $N=3\times 10^4$ bursts detected using this simple metric.
	
	\section{Investigating H-reionization using FRB redshifts}
	\label{frb-z}
	
\begin{figure}
		\centering
		\includegraphics[width = .4\textwidth]{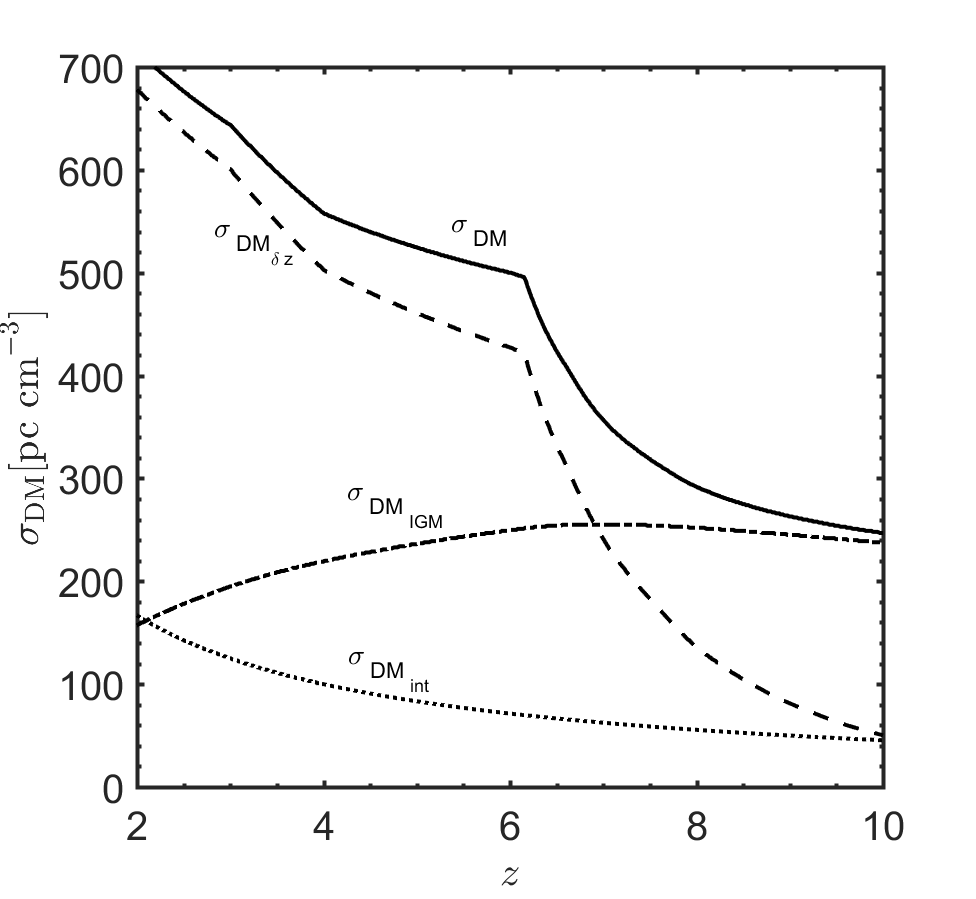}\\
		\includegraphics[width = .4\textwidth]{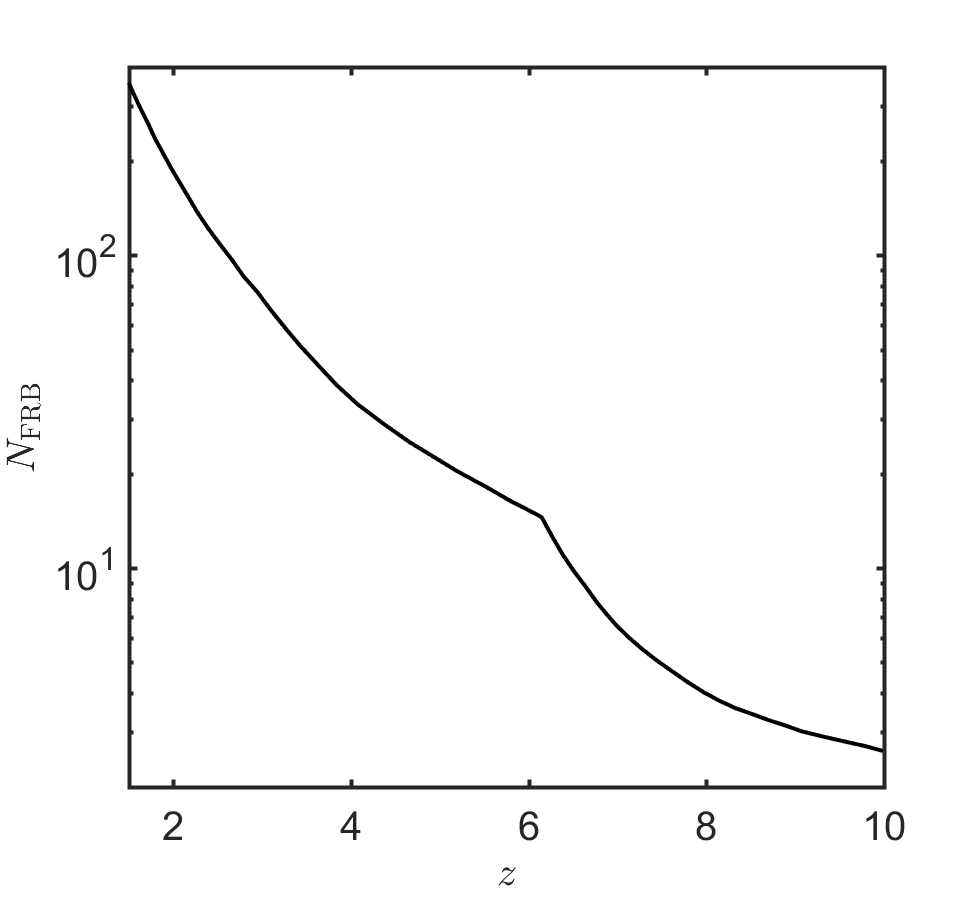}
		\vskip 0.1cm
		\caption{The upper panel shows contributions to $\sigma_{_{DM}}$ from
			IGM electron density fluctuations (dash-dot curve marked with 
			$\sigma_{_{DM_{\rm IGM}}}$, using the $\xi_{\rm e,o}(z)$ reionization history), host galaxy+CGM from FIRE simulations (dots, labeled
			$\sigma_{DM_{\rm int}}$), due to error associated with FRB host galaxy redshift measurements which is taken to be $0.5$ for this calculation, and due to the finite redshift bin-width which we took to be 1.0 (dash-curve, marked as $\sigma_{DM_{\delta z}}$, see Eq. \ref{sig-DMz} and calculated for $\xi_{\rm e,o}(z)$); we have taken $\sigma_{_{DM_{\rm IGM}}}(z) = 0.13\,\mbox{DM}_{\rm IGM}/(1+z)^{1/2}$ as per the \protect\cite{Jaroszynski2019} analysis of cosmological simulations. The square-root of the
			squared sum of these contributions is $\sigma_{_{DM}}$ (solid line). Note that $\sigma_{_{DM_{\rm int}}}$ decreases with redshift as $(1+z)^{-1}$ in the observer frame. The lower panel shows the number of FRBs for which redshift should be measured with an accuracy of $\Delta z = 0.5$ in order to determine $\langle\mbox{DM}_{\rm IGM}\rangle$ at the average redshift of these 
			bursts with an accuracy of 2.5\%.}
		\label{fig:sig_DM}
	\end{figure}

	DSA-1000 operating between 0.7 GHz \& 2 GHz, when online in a couple of years, would localize FRBs to within a few arc-seconds \citep{Hallinan2019}. A rough
	estimate of the FRB redshift, with error $\Delta z\sim 1$, can be obtained from the DM of the burst\footnote{Fluctuations in electron column density in the IGM, the FRB host galaxy and its circum-galactic medium, cause the DM of FRBs at a fixed redshift, but along different lines of sights, to fluctuate by $\sim 300$ pc cm$^{-3}$ (see Fig. \ref{fig:sig_DM}) for $z \gae 4$, which is about 10\% of the total DM at these redshifts. This limits the accuracy of determining burst redshift from DM to $\Delta z\sim 1$.} thereby making it possible to tag high-$z$ bursts ($z>6$) for follow up observations and redshift measurements; the error in redshift determination from DM increases during the reionization epoch when the neutral-H fraction is larger than a few 10\%. A small number of high-$z$ ($z>6$) FRBs are likely to be identified in follow up optical and IR observations as the number of galaxies per square-arcsecond at $z\lae 8$ is estimated to be 0.4 (e.g. \citealt{Finkelstein2016}), and for these bursts spectroscopy of the host galaxies could yield redshifts with an error of $\sim 10$\% by JWST. The follow up optical/IR observations are essential for establishing that the targeted FRB is not at a redshift much smaller than implied by its DM\footnote{If the FRB host galaxy or its CGM were to make an unusually large contribution to the observed DM, then in that case the redshift estimated from the DM would be way off. Follow up optical/IR observations can eliminate these bursts from further consideration for the purpose of investigating hydrogen reionization epoch.}, i.e. there is no galaxy nearby within the few arcsec FRB localization circle in the sky, and therefore the burst belongs in the sample for exploring the reionization epoch.
	
	The measurement of redshifts and DMs for a small sub-sample of FRBs 
	would be very useful for determining how the average electron density (per cc) 
	in the IGM varies with redshift, $n_{\rm IGM}(z)$, during the reionization epoch.
	This is facilitated by the fact that the contributions to the DM
	from the FRB host galaxy and CGM is relatively small (see \S\ref{fire-DM}),
	and the contribution from our galaxy can be subtracted reasonably well.
	What's more, determining $n_{\rm e}(z)$ is more reliable from $\mbox{DM}(z)$ 
	than it is using the Thomson scattering optical depth $\tau_{\rm T}(z)$, as  
	the latter quantity depends on the integral of electron density weighted by 
	$(1+z)^2$ -- see Eq. \ref{eq:tauT} -- whereas the DM integral has a weight 
	factor of $(1+z)$.
	
	Let us consider that redshifts of $N_i$ FRBs are measured -- from their DMs, photometrically or spectroscopically -- to be between $z_i\pm \delta_{1i}$ (in other words, $2\delta_{1i}$ is the width of the bin around $z_{i}$). The contributions to the DM of a burst at redshift $z$ from IGM is $\mbox{DM}_{\rm IGM}(z) \pm \sigma_{_{DM_{\rm IGM}}}(z)$, and the host galaxy and the CGM is $\mbox{DM}_{\rm int}(z) \pm \sigma_{_{DM_{\rm int}}}(z)$. Let us take the average error in redshift measurement at $z_i$ to be $\delta_{2i}$. The variance of FRB-DMs for a large sample of bursts due to their different redshifts, $z_i - \delta_{1i} \le z \le z_i + \delta_{1i}$, and error in redshift measurements ($\pm \delta_{2i}$), is given by
	\begin{equation}
	\sigma_{_{\rm DM_{\delta z}}} = {d\mbox{DM}_{\rm IGM}\over dz} \left[ \delta_{1i}^2 
	+ \delta_{2i}^2 \right]^{1/2}.
	\label{sig-DMz}
	\end{equation}
	Adding up the various contributions to the variance of FRB-DMs yields
	\begin{equation}
	\sigma_{_{\rm DM}}^2(z) = \sigma^2_{_{\rm DM_{\rm IGM}}}(z) + \sigma^2_{_{\rm DM_{\rm int}}}(z)
	+ \sigma^2_{_{\rm DM_{\delta z}}}.
	\end{equation}
	Both $\sigma_{_{\rm DM_{\delta z}}}$ and $\sigma_{_{\rm DM_{\rm IGM}}}$ depend on $\mbox{DM}_{\rm IGM}$ and therefore on the reionization history. However, the result that $\sigma_{_{\rm DM_{\delta z}}}$ should drop rapidly at $z\gtrsim 6$ (and that $\sigma_{_{DM_{\rm IGM}}}$ should then dominate the $\rm{DM}$ error) is generic, as it stems from the shallow evolution of $\rm{DM}(z)$ at large $z$ (see e.g. Fig. \ref{fig:dm-z}). Furthermore, to a first approximation, the relative error, $\sigma_{_{DM_{\rm IGM}}}/DM= 0.13/(1+z)^{0.5}$ is itself independent of the electron fraction \citep{Jaroszynski2019}.
	
	The average DM for a sample of $N_i$ FRBs can be written as
	\begin{equation}
	\begin{split}
	\langle\mbox{DM}(z_i)\rangle & \equiv {1\over N_i} \sum_{j=1}^{N_i}\mbox{DM}_j \\ & = 
	\langle\mbox{DM}_{\rm IGM}(z_i)\rangle + \langle\mbox{DM}_{\rm int}(z_i)\rangle 
	\pm {\sigma_{_{\rm DM}}(z_i)\over N_i^{1/2}},
	\end{split}
	\end{equation}
	where $\langle\mbox{DM}_{\rm IGM}(z_i)\rangle$ is the mean electron column density 
	in the IGM up to redshift $z_i$, and $\langle\mbox{DM}_{\rm int}\rangle$ is the 
	average contribution to the DM from the FRB host galaxy and its CGM.
	We see in Fig. \ref{fig:dPdDM} that the contributions of the FRB host galaxy and the CGM to the total DM of an FRB, $\langle\mbox{DM}_{\rm int}\rangle$, is of order 300 in the rest frame of the burst, which is more or less independent of the FRB redshift. The $\langle\mbox{DM}_{\rm int}\rangle$ in the observer frame is smaller by a factor $(1+z)$, and therefore for bursts at redshifts larger than 5 -- the domain of exploration in this work -- $\langle\mbox{DM}_{\rm int}(z_i)\rangle < 10^2\mbox{pc cm}^{-3}$ or less than 2\% of the total DM. The value of $ \sigma_{_{\rm DM}}(z)$ is plotted in Fig. \ref{fig:sig_DM}.
	We see that $\sigma_{_{\rm DM}} \lae 500\mbox{pc cm}^{-3}$ for $z\gae 5$. Therefore, to determine $\langle \mbox{DM}_{\rm IGM}\rangle$ at $z=5.5$ with an accuracy of 2.5\%, one needs to find $\sim20$ FRBs within a redshift bin of width 1.0 centered at 5.5 and the error in their redshift measurements of $\lae 0.5$. The lower panel of Fig. \ref{fig:sig_DM} shows the number of FRBs needed, in a redshift bin of width 1.0, for measuring $\langle \mbox{DM}_{\rm IGM}\rangle$ with 2.5\% accuracy at different redshifts. The average electron density in the IGM between $z_i$ and $z_{i+1}$ can be determined to $\sim 4$\% accuracy from the difference $\langle \mbox{DM}_{\rm IGM}(z_{i+1})\rangle - \langle \mbox{DM}_{\rm IGM}(z_i)\rangle$. One needs to measure redshifts of $\sim 40$ FRBs between 6 \& 10 to obtain $\sim 4$\% accuracy for $\langle n_{\rm e}\rangle$ at four distinct $z$ (Fig. \ref{fig:sig_DM}).
	
	The fluctuation in electron density associated with randomly distributed
	ionized bubbles can be obtained from the excess in the variance for DM
	at $z>6$ compared with the theoretically expected value shown in 
	Fig. \ref{fig:sig_DM}.
	
	\begin{figure}
		\centering
		\includegraphics[width = .35\textwidth]{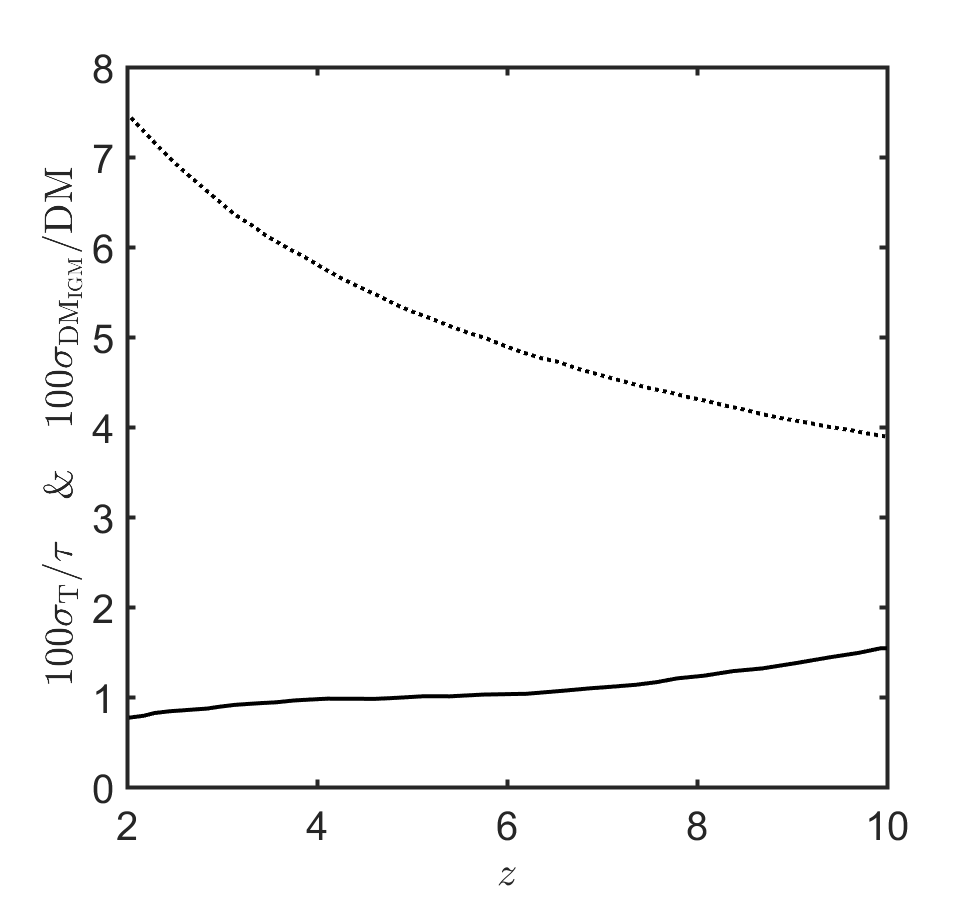}
		\vskip -0.2cm
		\caption{ Shown here are $\sigma_{_\tau}/\tau_{\rm T}$ calculated using Eq. 
			\ref{sig-tau} (solid line), and $\sigma_{_{DM_{\rm IGM}}}/DM= 0.13/(1+z)^{0.5}$ (dotted line); 
			both are these are \% error associated with the fluctuations in the 
			electron density of the IGM.
		}
		\label{fig:sig_tau}
	\end{figure} 
	
	Another quantity of interest is the Thompson optical depth as a function of
	$z$, $\tau_{\rm T}(z)$. The average value of $\tau_{\rm T}(z)$ can be easily determined from 
	$\langle \mbox{DM}_{\rm IGM}(z)\rangle$ as both of these are integrals of $\bar{n}_e(z)$
	with slightly different z-weight factors. 
	
	The variance of the optical depth, $\sigma^2_{_{\tau}}$, can be expressed
	in terms of $\sigma_{_{DM_{\rm IGM}}}$ as follows. From the definition of
	dispersion measure
	\begin{equation}
	\mbox{DM}_{\rm IGM}(z) = \int_0^{r(z)} dr_1 {n_{e}(r_1)\over (1+z_1)^2 } 
	\end{equation}
	it follows that
	\begin{equation}
	\sigma^2_{_{DM_{\rm IGM}}}(z) = \bar{n}_{e0}^2 \int dr_1 dr_2 \, (1+z_1) (1+z_2)
	\langle \delta_e(r_1) \delta_e(r_2) \rangle,
	\end{equation}
	where $\delta_e(r) \equiv [n_{\rm e}(r) - \bar{n}_e(r)]/\bar{n}_e$. We can write
	$ \langle \delta_e(r1) \delta_e(r_2) \rangle = \chi(z, |r_1-r_2|)$; $z =
	(z_1+z_2)/2$. The above integral can then be rewritten as
	\begin{equation}
	\begin{split}
	\sigma^2_{_{DM_{\rm IGM}}}(z) &= c\bar{n}_{e0}^2 \int dr (1+z)^2 \chi_1(z) \\
	&= c \bar{n}_{e0}^2 \int_0^z dz' {(1+z')^2 \chi_1(z') \over H(z')},
	\end{split}
	\label{sig-DM2}
	\end{equation}
	where $\chi_1(z) \equiv \int d(\delta r) \chi(z, \delta r)$. Similarly,
	starting with the expression for $\tau_{\rm T}(z)$
	\begin{equation}
	\tau_{\rm T}(z) = \sigma_T \int_0^{r(z)} dr_1 {n_{e}(r_1)\over (1+z_1) } 
	\end{equation}
	we arrive at
	\begin{equation}
	\sigma^2_{_{\tau}}(z) = c\sigma_T^2 \bar{n}_{e0}^2 \int_0^z dz' {(1+z')^4 
		\chi_1(z') \over H(z')}.
	\end{equation}
	We can replace $\chi_1(z)$ in terms of $d\sigma_{_{\rm DM_{\rm IGM}}}/dz$ using
	eq. (\ref{sig-DM2}), and rewrite the variance of $\tau_{\rm T}$ as
	\begin{equation}
	\begin{split}
	\sigma^2_{_{\tau}}(z) & = \sigma_T^2 \int_0^z dz' (1+z')^2 
	{ d \sigma_{_{\rm DM_{\rm IGM}}} \over dz'} \\
	&\! =\! \sigma_T^2 \left[ (1\!+\!z)^2\sigma_{_{\rm DM_{\rm IGM}}}(z)\!-\!2\int_0^z dz' 
	(1\!+\!z') \sigma_{_{\rm DM_{\rm IGM}}}(z') \right].
	\label{sig-tau}
	\end{split}
	\end{equation}
	Because of the strong redshift weighting of the above integral, 
	$\sigma_{_{\tau}}/\tau_{\rm T} < \sigma_{_{\mbox{DM}_{\rm IGM}}}/\mbox{DM}$ as 
	$d \sigma_{_{\rm DM_{\rm IGM}}}/dz$ is small at $z \gae 4$ (Fig. \ref{fig:sig_DM}).
	
	\section{Discussion and Conclusions}
	
	We have investigated in this work whether FRBs can be used to explore the hydrogen reionization epoch of the universe. The answer we find is affirmative. We have developed three different methods for probing the reionization epoch using FRBs, which are summarized below.

	\begin{enumerate}
		\item {\bf ${\rm DM}_{\rm max}$}: Because of the finite electron column 
		density between us and redshift $\sim 15$ when the universe was mostly neutral, the maximum DM of FRBs is ${\rm DM}_{\rm max} \sim 6{\rm x}10^3$ pc cm$^{-3}$. An accurate measurement of ${\rm DM}_{\rm max}$ provides a good way to constrain the gross features of H-reionization (mean $z$, and interval over which ionization took place). We have shown in \S\ref{sec:dmmax} that if ${\rm DM}_{\rm max}$ can be measured to a precision better than $500\,\mbox{pc cm}^{-3}$, then that would be more constraining of the reionization history, $\xi_e(z)$, than the Thomson optical depth measurement of Planck \citep{Planck2020}. This method does not require any knowledge of the FRB luminosity function and its evolution (as long as there are FRBs at $z>6$). Furthermore, completeness of the survey to some specific-fluence limit is not required for this method to work. However, one caveat is that in a rare case, a low-$z$ host-galaxy of an FRB ($z \lae 3$) might have $\mbox{DM}_{\rm int} \gae 3{\rm x}10^3$ pc cm$^{-3}$, which can compromise the use of $DM_{\rm max}$ to investigate the H-reionization history. To remove this bias, we suggest follow up optical observations of FRBs with total observed DM larger than $\sim 5500$ pc cm$^{-3}$ to ensure that the burst is not located in a low-$z$ galaxy with abnormally large ${\rm DM}_{\rm int}$.
		
		\item {\bf $d\dot{N}_{\rm FRB}/dDM$}: This method does not require redshift measurements for FRBs. It relies on the fact that the rate of FRBs per unit DM, $d\dot{N}_{\rm FRB}/d\mbox{DM}$, depends on the reionization history of the Universe and the FRB spectral energy distribution function. With the knowledge of the energy distribution function -- determined from a sample of FRBs with known redshifts -- one can determine the reionization history from $d\dot{N}_{\rm FRB}/d\mbox{DM}$. As an example, we considered the possibility that the FRB distribution follows the star formation rate, and found that, according to the Monte Carlo simulations' result, one needs to observe $\sim 3\times 10^4$ bursts to be able to determine any deviation of the reionization history from the one suggested in \cite{Robertson2015}. This calculation made use of FIRE simulations to estimate the contributions of FRB host galaxies and the CGM to the DM, and the analysis of \cite{Jaroszynski2019} of large scale {\it Illustris} simulations for the IGM DM distribution. The major sources of uncertainty associated with this approach are the FRB energy distribution function and its (potential) redshift evolution, as well as the width of the distribution of DM$_{\rm int}$, which is the contribution to the observed DM value from the FRB host-galaxy and the CGM. A related, but possibly more robust, method we have proposed is to measure the ratio of number of bursts in different DM bins (for instance, $5000-6000$ and $6000-7000\mbox{ pc cm}^{-3}$). The ratio is found to be strongly 
		dependent on the hydrogen reionization history.
		
		\item {\bf Redshifts of a small sample of $z>6$ FRBs:}  
		We have shown in \S\ref{frb-z} that a relatively small number of FRBs, of order 40, needs to be identified during the reionization epoch and their redshift measured with an accuracy of 5-10\%, to determine average ionization fraction in four redshift bins in the z-interval 6-9 to within $\sim 4$\%. This is because the errors arising from the density fluctuations in the IGM and the FRB host galaxy \& CGM are relatively small at high z; the latter contributions (galaxy and CGM) are suppressed by a factor (1+z). This method requires no knowledge of the FRB energy distribution function and its redshift evolution. Furthermore, it makes no assumption regarding the completeness of the survey.
	\end{enumerate}
	
	{\bf Data availability} The data produced in this study will be shared on reasonable request to the authors.
	
	\section{acknowledgments}
	We thank Mukul Bhattacharya, Mike Boylan-Kolchin, Liam Connor, Daniel Eisenstein, Steve Finkelstein, Eric Linder, Wenbin Lu, Rob Robinson and Paul Shapiro for useful conversations and input. In particular we thank Mukul Bhattacharya, Mike Boylan-Kolchin, Eric Linder and Wenbin Lu for reading the draft and providing numerous suggestions that helped improve the presentation significantly.
	The research of PB was funded by the Gordon and Betty Moore Foundation 
	through Grant GBMF5076. This work has been funded in part by an NSF 
	grant AST-2009619 and NSF grant AST-1715070.


\begin{thebibliography}{}
		\makeatletter
		\relax
		\def\mn@urlcharsother{\let\do\@makeother \do\$\do\&\do\#\do\^\do\_\do\%\do\~}
		\def\mn@doi{\begingroup\mn@urlcharsother \@ifnextchar [ {\mn@doi@}
			{\mn@doi@[]}}
		\def\mn@doi@[#1]#2{\def\@tempa{#1}\ifx\@tempa\@empty \href
			{http://dx.doi.org/#2} {doi:#2}\else \href {http://dx.doi.org/#2} {#1}\fi
			\endgroup}
		\def\mn@eprint#1#2{\mn@eprint@#1:#2::\@nil}
		\def\mn@eprint@arXiv#1{\href {http://arxiv.org/abs/#1} {{\tt arXiv:#1}}}
		\def\mn@eprint@dblp#1{\href {http://dblp.uni-trier.de/rec/bibtex/#1.xml}
			{dblp:#1}}
		\def\mn@eprint@#1:#2:#3:#4\@nil{\def\@tempa {#1}\def\@tempb {#2}\def\@tempc
			{#3}\ifx \@tempc \@empty \let \@tempc \@tempb \let \@tempb \@tempa \fi \ifx
			\@tempb \@empty \def\@tempb {arXiv}\fi \@ifundefined
			{mn@eprint@\@tempb}{\@tempb:\@tempc}{\expandafter \expandafter \csname
				mn@eprint@\@tempb\endcsname \expandafter{\@tempc}}}
		
		\bibitem[\protect\citeauthoryear{{Bannister} et~al.,}{{Bannister}
			et~al.}{2017}]{Bannister2017}
		{Bannister} K.~W.,  et~al., 2017, \mn@doi [\apjl] {10.3847/2041-8213/aa71ff},
		\href {https://ui.adsabs.harvard.edu/abs/2017ApJ...841L..12B} {841, L12}
		
		\bibitem[\protect\citeauthoryear{{Bannister} et~al.,}{{Bannister}
			et~al.}{2019}]{Bannister+19}
		{Bannister} K.~W.,  et~al., 2019, \mn@doi [Science] {10.1126/science.aaw5903},
		\href {https://ui.adsabs.harvard.edu/abs/2019Sci...365..565B} {365, 565}
		
		\bibitem[\protect\citeauthoryear{{Behroozi}, {Wechsler}, {Hearin}  \&
			{Conroy}}{{Behroozi} et~al.}{2019}]{Behroozi2019}
		{Behroozi} P.,  {Wechsler} R.~H.,  {Hearin} A.~P.,   {Conroy} C.,  2019,
		\mn@doi [\mnras] {10.1093/mnras/stz1182}, \href
		{https://ui.adsabs.harvard.edu/abs/2019MNRAS.488.3143B} {488, 3143}
		
		\bibitem[\protect\citeauthoryear{{Beniamini}, {Hotokezaka}, {van der Horst}  \&
			{Kouveliotou}}{{Beniamini} et~al.}{2019}]{Beniamini2019}
		{Beniamini} P.,  {Hotokezaka} K.,  {van der Horst} A.,   {Kouveliotou} C.,
		2019, \mn@doi [\mnras] {10.1093/mnras/stz1391}, \href
		{https://ui.adsabs.harvard.edu/abs/2019MNRAS.487.1426B} {487, 1426}
		
		\bibitem[\protect\citeauthoryear{{Beniamini}, {Wadiasingh}  \&
			{Metzger}}{{Beniamini} et~al.}{2020}]{Beniamini+20}
		{Beniamini} P.,  {Wadiasingh} Z.,   {Metzger} B.~D.,  2020, \mn@doi [\mnras]
		{10.1093/mnras/staa1783}, \href
		{https://ui.adsabs.harvard.edu/abs/2020MNRAS.496.3390B} {496, 3390}
		
		\bibitem[\protect\citeauthoryear{{Bhattacharya}, {Kumar}  \&
			{Linder}}{{Bhattacharya} et~al.}{2020}]{Bhattacharya2020}
		{Bhattacharya} M.,  {Kumar} P.,   {Linder} E.~V.,  2020, arXiv e-prints, \href
		{https://ui.adsabs.harvard.edu/abs/2020arXiv201014530B} {p. arXiv:2010.14530}
		
		\bibitem[\protect\citeauthoryear{{Bochenek}, {Ravi}, {Belov}, {Hallinan},
			{Kocz}, {Kulkarni}  \& {McKenna}}{{Bochenek} et~al.}{2020}]{STARE2020}
		{Bochenek} C.~D.,  {Ravi} V.,  {Belov} K.~V.,  {Hallinan} G.,  {Kocz} J.,
		{Kulkarni} S.~R.,   {McKenna} D.~L.,  2020, \mn@doi [\nat]
		{10.1038/s41586-020-2872-x}, \href
		{https://ui.adsabs.harvard.edu/abs/2020Natur.587...59B} {587, 59}
		
		\bibitem[\protect\citeauthoryear{{CHIME/FRB Collaboration} et~al.,}{{CHIME/FRB
				Collaboration} et~al.}{2019a}]{CHIME2019}
		{CHIME/FRB Collaboration} et~al., 2019a, \mn@doi [\nat]
		{10.1038/s41586-018-0867-7}, \href
		{https://ui.adsabs.harvard.edu/abs/2019Natur.566..230C} {566, 230}
		
		\bibitem[\protect\citeauthoryear{{CHIME/FRB Collaboration} et~al.,}{{CHIME/FRB
				Collaboration} et~al.}{2019b}]{CHIME2019b}
		{CHIME/FRB Collaboration} et~al., 2019b, \mn@doi [\nat]
		{10.1038/s41586-018-0864-x}, \href
		{https://ui.adsabs.harvard.edu/abs/2019Natur.566..235C} {566, 235}
		
		\bibitem[\protect\citeauthoryear{{Caleb}, {Flynn}  \& {Stappers}}{{Caleb}
			et~al.}{2019}]{Caleb2019}
		{Caleb} M.,  {Flynn} C.,   {Stappers} B.~W.,  2019, \mn@doi [\mnras]
		{10.1093/mnras/stz571}, \href
		{https://ui.adsabs.harvard.edu/abs/2019MNRAS.485.2281C} {485, 2281}
		
		\bibitem[\protect\citeauthoryear{{Chabrier}}{{Chabrier}}{2003}]{Chabrier2003}
		{Chabrier} G.,  2003, \mn@doi [\pasp] {10.1086/376392}, \href
		{https://ui.adsabs.harvard.edu/abs/2003PASP..115..763C} {115, 763}
		
		\bibitem[\protect\citeauthoryear{{Chatterjee} et~al.,}{{Chatterjee}
			et~al.}{2017}]{Chatterjee+17}
		{Chatterjee} S.,  et~al., 2017, \mn@doi [\nat] {10.1038/nature20797}, \href
		{https://ui.adsabs.harvard.edu/abs/2017Natur.541...58C} {541, 58}
		
		\bibitem[\protect\citeauthoryear{{Connor}, {Miller}  \& {Gardenier}}{{Connor}
			et~al.}{2020}]{Connor2020}
		{Connor} L.,  {Miller} M.~C.,   {Gardenier} D.~W.,  2020, \mn@doi [\mnras]
		{10.1093/mnras/staa2074}, \href
		{https://ui.adsabs.harvard.edu/abs/2020MNRAS.497.3076C} {497, 3076}
		
		\bibitem[\protect\citeauthoryear{{Cordes} \& {Lazio}}{{Cordes} \&
			{Lazio}}{2002}]{CL2002}
		{Cordes} J.~M.,  {Lazio} T.~J.~W.,  2002, arXiv e-prints, \href
		{https://ui.adsabs.harvard.edu/abs/2002astro.ph..7156C} {pp
			astro--ph/0207156}
		
		\bibitem[\protect\citeauthoryear{{Dayal} \& {Ferrara}}{{Dayal} \&
			{Ferrara}}{2018}]{Dayal2018}
		{Dayal} P.,  {Ferrara} A.,  2018, \mn@doi [\physrep]
		{10.1016/j.physrep.2018.10.002}, \href
		{https://ui.adsabs.harvard.edu/abs/2018PhR...780....1D} {780, 1}
		
		\bibitem[\protect\citeauthoryear{{Deng} \& {Zhang}}{{Deng} \&
			{Zhang}}{2014}]{Deng2014}
		{Deng} W.,  {Zhang} B.,  2014, \mn@doi [\apjl] {10.1088/2041-8205/783/2/L35},
		\href {https://ui.adsabs.harvard.edu/abs/2014ApJ...783L..35D} {783, L35}
		
		\bibitem[\protect\citeauthoryear{{Eide}, {Ciardi}, {Graziani}, {Busch}, {Feng}
			\& {Di Matteo}}{{Eide} et~al.}{2020}]{Eide2020}
		{Eide} M.~B.,  {Ciardi} B.,  {Graziani} L.,  {Busch} P.,  {Feng} Y.,   {Di
			Matteo} T.,  2020, \mn@doi [\mnras] {10.1093/mnras/staa2774}, \href
		{https://ui.adsabs.harvard.edu/abs/2020MNRAS.498.6083E} {498, 6083}
		
		\bibitem[\protect\citeauthoryear{{Fan} et~al.,}{{Fan} et~al.}{2006}]{Fan2006}
		{Fan} X.,  et~al., 2006, \mn@doi [\aj] {10.1086/504836}, \href
		{https://ui.adsabs.harvard.edu/abs/2006AJ....132..117F} {132, 117}
		
		\bibitem[\protect\citeauthoryear{{Farah} et~al.,}{{Farah}
			et~al.}{2018}]{Farah2018}
		{Farah} W.,  et~al., 2018, \mn@doi [\mnras] {10.1093/mnras/sty1122}, \href
		{https://ui.adsabs.harvard.edu/abs/2018MNRAS.478.1209F} {478, 1209}
		
		\bibitem[\protect\citeauthoryear{{Finkelstein}}{{Finkelstein}}{2016}]{Finkelstein2016}
		{Finkelstein} S.~L.,  2016, \mn@doi [\pasa] {10.1017/pasa.2016.26}, \href
		{https://ui.adsabs.harvard.edu/abs/2016PASA...33...37F} {33, e037}
		
		\bibitem[\protect\citeauthoryear{{Finkelstein} et~al.,}{{Finkelstein}
			et~al.}{2019}]{Finkelstein2019}
		{Finkelstein} S.~L.,  et~al., 2019, \mn@doi [\apj] {10.3847/1538-4357/ab1ea8},
		\href {https://ui.adsabs.harvard.edu/abs/2019ApJ...879...36F} {879, 36}
		
		\bibitem[\protect\citeauthoryear{{Gajjar} et~al.,}{{Gajjar}
			et~al.}{2018}]{Gajjar2018}
		{Gajjar} V.,  et~al., 2018, \mn@doi [\apj] {10.3847/1538-4357/aad005}, \href
		{https://ui.adsabs.harvard.edu/abs/2018ApJ...863....2G} {863, 2}
		
		\bibitem[\protect\citeauthoryear{{Gunn} \& {Peterson}}{{Gunn} \&
			{Peterson}}{1965}]{GP1965}
		{Gunn} J.~E.,  {Peterson} B.~A.,  1965, \mn@doi [\apj] {10.1086/148444}, \href
		{https://ui.adsabs.harvard.edu/abs/1965ApJ...142.1633G} {142, 1633}
		
		\bibitem[\protect\citeauthoryear{{Hallinan} et~al.,}{{Hallinan}
			et~al.}{2019}]{Hallinan2019}
		{Hallinan} G.,  et~al., 2019, in Bulletin of the American Astronomical Society.
		p.~255 (\mn@eprint {arXiv} {1907.07648})
		
		\bibitem[\protect\citeauthoryear{{Hashimoto} et~al.,}{{Hashimoto}
			et~al.}{2020}]{Hashimoto2020}
		{Hashimoto} T.,  et~al., 2020, \mn@doi [\mnras] {10.1093/mnras/staa2490}, \href
		{https://ui.adsabs.harvard.edu/abs/2020MNRAS.498.3927H} {498, 3927}
		
		\bibitem[\protect\citeauthoryear{{Hoag} et~al.,}{{Hoag}
			et~al.}{2019}]{Hoag2019}
		{Hoag} A.,  et~al., 2019, \mn@doi [\apj] {10.3847/1538-4357/ab1de7}, \href
		{https://ui.adsabs.harvard.edu/abs/2019ApJ...878...12H} {878, 12}
		
		\bibitem[\protect\citeauthoryear{{Inoue}}{{Inoue}}{2004}]{Inoue2004}
		{Inoue} S.,  2004, \mn@doi [\mnras] {10.1111/j.1365-2966.2004.07359.x}, \href
		{https://ui.adsabs.harvard.edu/abs/2004MNRAS.348..999I} {348, 999}
		
		\bibitem[\protect\citeauthoryear{{Ioka}}{{Ioka}}{2003}]{Ioka2003}
		{Ioka} K.,  2003, \mn@doi [\apjl] {10.1086/380598}, \href
		{https://ui.adsabs.harvard.edu/abs/2003ApJ...598L..79I} {598, L79}
		
		\bibitem[\protect\citeauthoryear{{Jaroszynski}}{{Jaroszynski}}{2019}]{Jaroszynski2019}
		{Jaroszynski} M.,  2019, \mn@doi [\mnras] {10.1093/mnras/sty3529}, \href
		{https://ui.adsabs.harvard.edu/abs/2019MNRAS.484.1637J} {484, 1637}
		
		\bibitem[\protect\citeauthoryear{{Jaroszy{\'n}ski}}{{Jaroszy{\'n}ski}}{2020}]{Jaroszynski2020}
		{Jaroszy{\'n}ski} M.,  2020, \mn@doi [\actaa] {10.32023/0001-5237/70.2.1},
		\href {https://ui.adsabs.harvard.edu/abs/2020AcA....70...87J} {70, 87}
		
		\bibitem[\protect\citeauthoryear{{Kocz} et~al.,}{{Kocz}
			et~al.}{2019}]{Kocz2019}
		{Kocz} J.,  et~al., 2019, \mn@doi [\mnras] {10.1093/mnras/stz2219}, \href
		{https://ui.adsabs.harvard.edu/abs/2019MNRAS.489..919K} {489, 919}
		
		\bibitem[\protect\citeauthoryear{{Kroupa}}{{Kroupa}}{2001}]{Kroupa2001}
		{Kroupa} P.,  2001, \mn@doi [\mnras] {10.1046/j.1365-8711.2001.04022.x}, \href
		{https://ui.adsabs.harvard.edu/abs/2001MNRAS.322..231K} {322, 231}
		
		\bibitem[\protect\citeauthoryear{{Kumar} \& {Zhang}}{{Kumar} \&
			{Zhang}}{2015}]{KZ2015}
		{Kumar} P.,  {Zhang} B.,  2015, \mn@doi [\physrep]
		{10.1016/j.physrep.2014.09.008}, \href
		{https://ui.adsabs.harvard.edu/abs/2015PhR...561....1K} {561, 1}
		
		\bibitem[\protect\citeauthoryear{{Law} et~al.,}{{Law} et~al.}{2017}]{Law+17}
		{Law} C.~J.,  et~al., 2017, \mn@doi [\apj] {10.3847/1538-4357/aa9700}, \href
		{https://ui.adsabs.harvard.edu/abs/2017ApJ...850...76L} {850, 76}
		
		\bibitem[\protect\citeauthoryear{{Linder}}{{Linder}}{2020}]{Linder2020}
		{Linder} E.~V.,  2020, \mn@doi [\prd] {10.1103/PhysRevD.101.103019}, \href
		{https://ui.adsabs.harvard.edu/abs/2020PhRvD.101j3019L} {101, 103019}
		
		\bibitem[\protect\citeauthoryear{{Lorimer}, {Bailes}, {McLaughlin}, {Narkevic}
			\& {Crawford}}{{Lorimer} et~al.}{2007}]{Lorimer+07}
		{Lorimer} D.~R.,  {Bailes} M.,  {McLaughlin} M.~A.,  {Narkevic} D.~J.,
		{Crawford} F.,  2007, \mn@doi [Science] {10.1126/science.1147532}, \href
		{https://ui.adsabs.harvard.edu/abs/2007Sci...318..777L} {318, 777}
		
		\bibitem[\protect\citeauthoryear{{Lu} \& {Piro}}{{Lu} \&
			{Piro}}{2019}]{LuPiro19}
		{Lu} W.,  {Piro} A.~L.,  2019, \mn@doi [\apj] {10.3847/1538-4357/ab3796}, \href
		{https://ui.adsabs.harvard.edu/abs/2019ApJ...883...40L} {883, 40}
		
		\bibitem[\protect\citeauthoryear{{Lu}, {Kumar}  \& {Zhang}}{{Lu}
			et~al.}{2020}]{LKZ2020}
		{Lu} W.,  {Kumar} P.,   {Zhang} B.,  2020, \mn@doi [\mnras]
		{10.1093/mnras/staa2450}, \href
		{https://ui.adsabs.harvard.edu/abs/2020MNRAS.498.1397L} {498, 1397}
		
		\bibitem[\protect\citeauthoryear{{Luo}, {Men}, {Lee}, {Wang}, {Lorimer}  \&
			{Zhang}}{{Luo} et~al.}{2020}]{Luo2020}
		{Luo} R.,  {Men} Y.,  {Lee} K.,  {Wang} W.,  {Lorimer} D.~R.,   {Zhang} B.,
		2020, \mn@doi [\mnras] {10.1093/mnras/staa704}, \href
		{https://ui.adsabs.harvard.edu/abs/2020MNRAS.494..665L} {494, 665}
		
		\bibitem[\protect\citeauthoryear{{Ma} et~al.,}{{Ma} et~al.}{2018}]{Ma2018}
		{Ma} X.,  et~al., 2018, \mn@doi [\mnras] {10.1093/mnras/sty1024}, \href
		{https://ui.adsabs.harvard.edu/abs/2018MNRAS.478.1694M} {478, 1694}
		
		\bibitem[\protect\citeauthoryear{{Ma}, {Quataert}, {Wetzel}, {Hopkins},
			{Faucher-Gigu{\`e}re}  \& {Kere{\v{s}}}}{{Ma} et~al.}{2020}]{Ma2020}
		{Ma} X.,  {Quataert} E.,  {Wetzel} A.,  {Hopkins} P.~F.,  {Faucher-Gigu{\`e}re}
		C.-A.,   {Kere{\v{s}}} D.,  2020, \mn@doi [\mnras] {10.1093/mnras/staa2404},
		\href {https://ui.adsabs.harvard.edu/abs/2020MNRAS.498.2001M} {498, 2001}
		
		\bibitem[\protect\citeauthoryear{{Macquart}, {Shannon}, {Bannister}, {James},
			{Ekers}  \& {Bunton}}{{Macquart} et~al.}{2019}]{Macquart2019}
		{Macquart} J.~P.,  {Shannon} R.~M.,  {Bannister} K.~W.,  {James} C.~W.,
		{Ekers} R.~D.,   {Bunton} J.~D.,  2019, \mn@doi [\apjl]
		{10.3847/2041-8213/ab03d6}, \href
		{https://ui.adsabs.harvard.edu/abs/2019ApJ...872L..19M} {872, L19}
		
		\bibitem[\protect\citeauthoryear{{Macquart} et~al.,}{{Macquart}
			et~al.}{2020}]{Macquart20}
		{Macquart} J.~P.,  et~al., 2020, \mn@doi [\nat] {10.1038/s41586-020-2300-2},
		\href {https://ui.adsabs.harvard.edu/abs/2020Natur.581..391M} {581, 391}
		
		\bibitem[\protect\citeauthoryear{{Madau} \& {Dickinson}}{{Madau} \&
			{Dickinson}}{2014}]{MD2014}
		{Madau} P.,  {Dickinson} M.,  2014, \mn@doi [\araa]
		{10.1146/annurev-astro-081811-125615}, \href
		{https://ui.adsabs.harvard.edu/abs/2014ARA&A..52..415M} {52, 415}
		
		\bibitem[\protect\citeauthoryear{{Marcote} et~al.,}{{Marcote}
			et~al.}{2017}]{Marcote2017}
		{Marcote} B.,  et~al., 2017, \mn@doi [\apjl] {10.3847/2041-8213/834/2/L8},
		\href {https://ui.adsabs.harvard.edu/abs/2017ApJ...834L...8M} {834, L8}
		
		\bibitem[\protect\citeauthoryear{{Margalit}, {Beniamini}, {Sridhar}  \&
			{Metzger}}{{Margalit} et~al.}{2020}]{MBSM2020}
		{Margalit} B.,  {Beniamini} P.,  {Sridhar} N.,   {Metzger} B.~D.,  2020,
		\mn@doi [\apjl] {10.3847/2041-8213/abac57}, \href
		{https://ui.adsabs.harvard.edu/abs/2020ApJ...899L..27M} {899, L27}
		
		\bibitem[\protect\citeauthoryear{{Mason}, {Treu}, {Dijkstra}, {Mesinger},
			{Trenti}, {Pentericci}, {de Barros}  \& {Vanzella}}{{Mason}
			et~al.}{2018}]{Mason2018}
		{Mason} C.~A.,  {Treu} T.,  {Dijkstra} M.,  {Mesinger} A.,  {Trenti} M.,
		{Pentericci} L.,  {de Barros} S.,   {Vanzella} E.,  2018, \mn@doi [\apj]
		{10.3847/1538-4357/aab0a7}, \href
		{https://ui.adsabs.harvard.edu/abs/2018ApJ...856....2M} {856, 2}
		
		\bibitem[\protect\citeauthoryear{{Mason} et~al.,}{{Mason}
			et~al.}{2019}]{Mason2019}
		{Mason} C.~A.,  et~al., 2019, \mn@doi [\mnras] {10.1093/mnras/stz632}, \href
		{https://ui.adsabs.harvard.edu/abs/2019MNRAS.485.3947M} {485, 3947}
		
		\bibitem[\protect\citeauthoryear{{Michilli} et~al.,}{{Michilli}
			et~al.}{2018}]{Michilli+18}
		{Michilli} D.,  et~al., 2018, \mn@doi [\nat] {10.1038/nature25149}, \href
		{https://ui.adsabs.harvard.edu/abs/2018Natur.553..182M} {553, 182}
		
		\bibitem[\protect\citeauthoryear{{Miller} \& {Scalo}}{{Miller} \&
			{Scalo}}{1979}]{MillerScalo79}
		{Miller} G.~E.,  {Scalo} J.~M.,  1979, \mn@doi [\apjs] {10.1086/190629}, \href
		{https://ui.adsabs.harvard.edu/abs/1979ApJS...41..513M} {41, 513}
		
		\bibitem[\protect\citeauthoryear{{Muno} et~al.,}{{Muno}
			et~al.}{2006}]{Muno2006}
		{Muno} M.~P.,  et~al., 2006, \mn@doi [\apjl] {10.1086/499776}, \href
		{https://ui.adsabs.harvard.edu/abs/2006ApJ...636L..41M} {636, L41}
		
		\bibitem[\protect\citeauthoryear{{Os{\l}owski} et~al.,}{{Os{\l}owski}
			et~al.}{2019}]{Oslowski2019}
		{Os{\l}owski} S.,  et~al., 2019, \mn@doi [\mnras] {10.1093/mnras/stz1751},
		\href {https://ui.adsabs.harvard.edu/abs/2019MNRAS.488..868O} {488, 868}
		
		\bibitem[\protect\citeauthoryear{{Paoletti}, {Hazra}, {Finelli}  \&
			{Smoot}}{{Paoletti} et~al.}{2020}]{Paoletti2020}
		{Paoletti} D.,  {Hazra} D.~K.,  {Finelli} F.,   {Smoot} G.~F.,  2020, \mn@doi
		[\jcap] {10.1088/1475-7516/2020/09/005}, \href
		{https://ui.adsabs.harvard.edu/abs/2020JCAP...09..005P} {2020, 005}
		
		\bibitem[\protect\citeauthoryear{{Petroff} et~al.,}{{Petroff}
			et~al.}{2016}]{Petroff2016}
		{Petroff} E.,  et~al., 2016, \mn@doi [\pasa] {10.1017/pasa.2016.35}, \href
		{https://ui.adsabs.harvard.edu/abs/2016PASA...33...45P} {33, e045}
		
		\bibitem[\protect\citeauthoryear{{Planck Collaboration} et~al.,}{{Planck
				Collaboration} et~al.}{2016}]{Planck2016}
		{Planck Collaboration} et~al., 2016, \mn@doi [\aap]
		{10.1051/0004-6361/201525830}, \href
		{https://ui.adsabs.harvard.edu/abs/2016A&A...594A..13P} {594, A13}
		
		\bibitem[\protect\citeauthoryear{{Planck Collaboration} et~al.,}{{Planck
				Collaboration} et~al.}{2020}]{Planck2020}
		{Planck Collaboration} et~al., 2020, \mn@doi [\aap]
		{10.1051/0004-6361/201936386}, \href
		{https://ui.adsabs.harvard.edu/abs/2020A&A...641A...5P} {641, A5}
		
		\bibitem[\protect\citeauthoryear{{Prochaska} et~al.,}{{Prochaska}
			et~al.}{2019}]{Prochaska+19}
		{Prochaska} J.~X.,  et~al., 2019, \mn@doi [Science] {10.1126/science.aay0073},
		\href {https://ui.adsabs.harvard.edu/abs/2019Sci...366..231P} {366, 231}
		
		\bibitem[\protect\citeauthoryear{{Ravi}}{{Ravi}}{2019a}]{Ravi2019b}
		{Ravi} V.,  2019a, \mn@doi [Nature Astronomy] {10.1038/s41550-019-0831-y},
		\href {https://ui.adsabs.harvard.edu/abs/2019NatAs...3..928R} {3, 928}
		
		\bibitem[\protect\citeauthoryear{{Ravi}}{{Ravi}}{2019b}]{Ravi2019}
		{Ravi} V.,  2019b, \mn@doi [\mnras] {10.1093/mnras/sty1551}, \href
		{https://ui.adsabs.harvard.edu/abs/2019MNRAS.482.1966R} {482, 1966}
		
		\bibitem[\protect\citeauthoryear{{Ravi} et~al.,}{{Ravi} et~al.}{2019}]{Ravi+19}
		{Ravi} V.,  et~al., 2019, \mn@doi [\nat] {10.1038/s41586-019-1389-7}, \href
		{https://ui.adsabs.harvard.edu/abs/2019Natur.572..352R} {572, 352}
		
		\bibitem[\protect\citeauthoryear{{Robertson}, {Ellis}, {Furlanetto}  \&
			{Dunlop}}{{Robertson} et~al.}{2015}]{Robertson2015}
		{Robertson} B.~E.,  {Ellis} R.~S.,  {Furlanetto} S.~R.,   {Dunlop} J.~S.,
		2015, \mn@doi [\apjl] {10.1088/2041-8205/802/2/L19}, \href
		{https://ui.adsabs.harvard.edu/abs/2015ApJ...802L..19R} {802, L19}
		
		\bibitem[\protect\citeauthoryear{{Shannon} et~al.,}{{Shannon}
			et~al.}{2018}]{Shannon2018}
		{Shannon} R.~M.,  et~al., 2018, \mn@doi [\nat] {10.1038/s41586-018-0588-y},
		\href {https://ui.adsabs.harvard.edu/abs/2018Natur.562..386S} {562, 386}
		
		\bibitem[\protect\citeauthoryear{{Spitler} et~al.,}{{Spitler}
			et~al.}{2014}]{Spitler2014}
		{Spitler} L.~G.,  et~al., 2014, \mn@doi [\apj] {10.1088/0004-637X/790/2/101},
		\href {https://ui.adsabs.harvard.edu/abs/2014ApJ...790..101S} {790, 101}
		
		\bibitem[\protect\citeauthoryear{{Tendulkar} et~al.,}{{Tendulkar}
			et~al.}{2017}]{Tendulkar+17}
		{Tendulkar} S.~P.,  et~al., 2017, \mn@doi [\apjl] {10.3847/2041-8213/834/2/L7},
		\href {https://ui.adsabs.harvard.edu/abs/2017ApJ...834L...7T} {834, L7}
		
		\bibitem[\protect\citeauthoryear{{The Chime/Frb Collaboration} Andersen
			et~al.,}{{The Chime/Frb Collaboration} et~al.}{2020}]{CHIME2020}
		{The Chime/Frb Collaboration} Andersen B.~C.,  et~al., 2020, \mn@doi [\nat]
		{10.1038/s41586-020-2863-y}, \href
		{https://ui.adsabs.harvard.edu/abs/2020Natur.587...54T} {587, 54}
		
		\bibitem[\protect\citeauthoryear{{Thompson} \& {Duncan}}{{Thompson} \&
			{Duncan}}{1993}]{TD1993}
		{Thompson} C.,  {Duncan} R.~C.,  1993, \mn@doi [\apj] {10.1086/172580}, \href
		{https://ui.adsabs.harvard.edu/abs/1993ApJ...408..194T} {408, 194}
		
		\bibitem[\protect\citeauthoryear{{Thornton} et~al.,}{{Thornton}
			et~al.}{2013}]{Thornton+13}
		{Thornton} D.,  et~al., 2013, \mn@doi [Science] {10.1126/science.1236789},
		\href {https://ui.adsabs.harvard.edu/abs/2013Sci...341...53T} {341, 53}
		
		\bibitem[\protect\citeauthoryear{{Wadiasingh}, {Beniamini}, {Timokhin},
			{Baring}, {van der Horst}, {Harding}  \& {Kazanas}}{{Wadiasingh}
			et~al.}{2020}]{Wadiasingh2020}
		{Wadiasingh} Z.,  {Beniamini} P.,  {Timokhin} A.,  {Baring} M.~G.,  {van der
			Horst} A.~J.,  {Harding} A.~K.,   {Kazanas} D.,  2020, \mn@doi [\apj]
		{10.3847/1538-4357/ab6d69}, \href
		{https://ui.adsabs.harvard.edu/abs/2020ApJ...891...82W} {891, 82}
		
		\bibitem[\protect\citeauthoryear{{Walters}, {Weltman}, {Gaensler}, {Ma}  \&
			{Witzemann}}{{Walters} et~al.}{2018}]{Walters2018}
		{Walters} A.,  {Weltman} A.,  {Gaensler} B.~M.,  {Ma} Y.-Z.,   {Witzemann} A.,
		2018, \mn@doi [\apj] {10.3847/1538-4357/aaaf6b}, \href
		{https://ui.adsabs.harvard.edu/abs/2018ApJ...856...65W} {856, 65}
		
		\bibitem[\protect\citeauthoryear{{Whitler}, {Mason}, {Ren}, {Dijkstra},
			{Mesinger}, {Pentericci}, {Trenti}  \& {Treu}}{{Whitler}
			et~al.}{2020}]{Whitler2020}
		{Whitler} L.~R.,  {Mason} C.~A.,  {Ren} K.,  {Dijkstra} M.,  {Mesinger} A.,
		{Pentericci} L.,  {Trenti} M.,   {Treu} T.,  2020, \mn@doi [\mnras]
		{10.1093/mnras/staa1178}, \href
		{https://ui.adsabs.harvard.edu/abs/2020MNRAS.495.3602W} {495, 3602}
		
		\bibitem[\protect\citeauthoryear{{Wucknitz}, {Spitler}  \& {Pen}}{{Wucknitz}
			et~al.}{2020}]{Wucknitz2020}
		{Wucknitz} O.,  {Spitler} L.~G.,   {Pen} U.~L.,  2020, arXiv e-prints, \href
		{https://ui.adsabs.harvard.edu/abs/2020arXiv200411643W} {p. arXiv:2004.11643}
		
		\bibitem[\protect\citeauthoryear{{Yang} \& {Zhang}}{{Yang} \&
			{Zhang}}{2017}]{YZ2017}
		{Yang} Y.-P.,  {Zhang} B.,  2017, \mn@doi [\apj] {10.3847/1538-4357/aa8721},
		\href {https://ui.adsabs.harvard.edu/abs/2017ApJ...847...22Y} {847, 22}
		
		\bibitem[\protect\citeauthoryear{{Yao}, {Manchester}  \& {Wang}}{{Yao}
			et~al.}{2017}]{Yao2017}
		{Yao} J.~M.,  {Manchester} R.~N.,   {Wang} N.,  2017, \mn@doi [\apj]
		{10.3847/1538-4357/835/1/29}, \href
		{https://ui.adsabs.harvard.edu/abs/2017ApJ...835...29Y} {835, 29}
		
		\bibitem[\protect\citeauthoryear{{Yu}}{{Yu}}{2014}]{Yu2014}
		{Yu} Y.-W.,  2014, \mn@doi [\apj] {10.1088/0004-637X/796/2/93}, \href
		{https://ui.adsabs.harvard.edu/abs/2014ApJ...796...93Y} {796, 93}
		
		\bibitem[\protect\citeauthoryear{{Zhang}}{{Zhang}}{2018}]{Zhang2018}
		{Zhang} B.,  2018, \mn@doi [\apjl] {10.3847/2041-8213/aae8e3}, \href
		{https://ui.adsabs.harvard.edu/abs/2018ApJ...867L..21Z} {867, L21}
		
		\makeatother
	\end{thebibliography}
\end{document}